# Situated Bayes - Feminist and Pluriversal Perspectives on Bayesian Knowledge

Juni Schindler, Department of Mathematics, Imperial College London, United Kingdom
Goda Klumbytė, Department of Participatory IT Design, University of Kassel, Germany
Matthew Fuller, Department of Media and Communications, Goldsmiths, University of London, United Kingdom

Juni Schindler and Goda Klumbytė contributed equally to this paper.

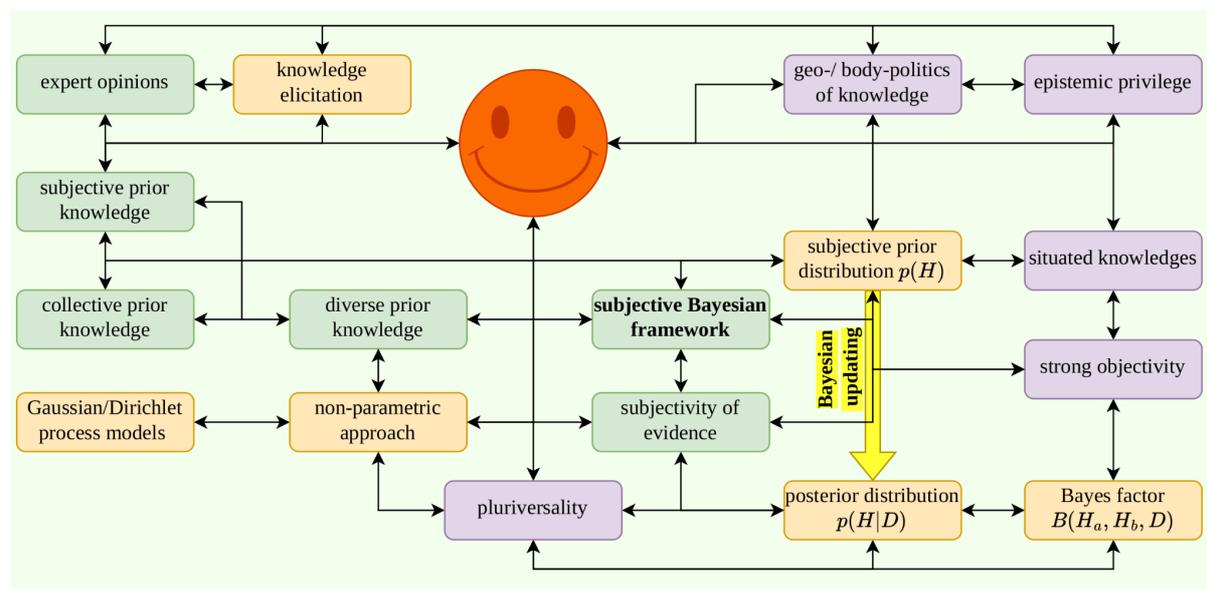

**Abstract:** This is the introduction and lead article to the Situated Bayes special issue of Computational Culture. The article introduces Bayes' Theorem and aspects of its contemporary uses, for instance in machine learning. A mathematical discussion is developed alongside a consideration of Bayes Theorem in relation to critical theories of knowledge, specifically the discussion of situated knowledge in feminist theories of science, pluriversal knowledge in decolonial theory, and critical approaches to mathematics. We discuss whether there are possible resonances between Bayesian mapping of multiple functions and the idea of the subjective on the one hand and these theoretical propositions on the other and propose further lines of enquiry for future research. In closing the introduction, the contributions to the special issue are briefly described.

# 1 Introduction

Today, versions of Bayes' Theorem have applications in all areas of social and natural sciences. The increase in probabilistic approaches entailed by the widespread application of methods of statistical computing have only increased the uptake of this approach. A significant body of work exists developing Bayesian approaches to epistemology drawing on broadly analytic methods.[1] This special issue aims to complement such work by drawing on different approaches to epistemology that take the political, aesthetic and social into greater account. At the same time, we recognise that there are significant distinctions to be drawn between the proposition-based and quantifying methods of statistical computation and what becomes the 'excluded middle' of philosophical or theoretical concepts.[2] This special issue thus is propositional and experimental, offering provisional attempts at bringing Bayesian approaches together with feminist and pluriversal perspectives.

**Bayesian inference**

Bayesian statistics understands probability as a measure of (subjective) confidence or belief. Subjective Bayesianism in turn is a statistical and philosophical approach to reasoning about conditional probabilities that conceptualises probability as an agent's subjective degree of belief and the grounds on which that belief might be based.[3] For example, we can express our knowledge about the values a certain parameter θ (often denoted by the lowercase Greek letter theta) can take with a probability distribution p(θ).[4] In Bayesian statistics, p(θ) is called the *prior distribution* and it can represent previously gained (subjective) experience. How does observing new data Y then change our belief in the distribution of θ? At the core of Bayesian data analysis is the process of *Bayesian inference* or *updating*, where p(θ) is updated to a *posterior distribution* p(θ | y) of θ, where (as Adrian Mackenzie discusses in his article for this issue) the vertical bar | denotes a conditional probability, resulting from new empirical observations Y using *Bayes' Theorem*:[5]

**Posterior**
Distribution of parameter $\theta$ after observing data $Y$.

**Likelihood**
Distribution of the data $Y$ given parameter $\theta$.

**Prior**
Distribution of parameter $\theta$ before observing data $Y$.

$$p(\theta|Y) = \frac{p(Y|\theta)\ p(\theta)}{p(Y)} \qquad (1)$$

**Evidence**
Distribution of the data $Y$ independent of the parameter $\theta$.

Here, the conditional probability p(Y | θ) denotes the distribution of the data Y under the condition that θ has certain values and the symbol | is used to express that conditionality. However, not all data might be available at the same time but arriving in a 'haphazard fashion',[6]

---

[1] Jan Sprenger and Stephan Hartmann, *Bayesian Philosophy of Science*, 1st ed. (Oxford University Press, 2019). Luc Bovens and Stephan Hartmann, *Bayesian Epistemology* (Oxford University Press, 2004).

[2] Gilles Deleuze, *Difference and Repetition*, trans. Paul Patton (Columbia University Press, 1995).

[3] Ian Hacking, *Logic of Statistical Inference* (Cambridge University Press, 2016). Scott M. Lynch and Bryce Bartlett, "Bayesian Statistics in Sociology: Past, Present, and Future", *Annual Review of Sociology* 45, no. 1 (2019): 47–68, https://doi.org/10.1146/annurev-soc-073018-022457. For a prosaic introduction to Subjective Bayes see: Armin Haas, "The Fabulous Five – A Bayesian Fairy Tale", in *Textures of the Anthropocene* (Cambridge, MA: MIT Press, 2015), 209–22.

[4] If Θ can only attain a finite number of values, p assigns each value a probability between 0 and 1 such that all probabilities sum to 1, and if Θ takes continuous values p becomes a density function. See Wikipedia for more details: https://en.wikipedia.org/wiki/Probability_distribution#Introduction.

[5] Andrew Gelman et al., *Bayesian Data Analysis*, 3rd ed. (CRC Press, 2014).

[6] Caitlin E. Buck and Bo Meson, "On Being a Good Bayesian", *World Archaeology* 47, no. 4 (2015): 567–84, https://doi.org/10.1080/00438243.2015.1053977.

such that the process of Bayesian updating can be repeated iteratively: After taking the posterior distribution p(θ | Y) as the new prior distribution, more recent (or other) data Y' enables again a revision of the subjective degrees of belief to produce a new posterior distribution p(θ | Y') by Bayesian updating. We refer to Section 3.2 for an in-depth treatment and illustration of Bayesian inference in the context of Bayesian linear regression.

**Situating Bayesian procedures**
Whilst in some respects, subjective Bayesianism describes or provides a diagrammatic form of a simplified notion of 'reason' (with the attendant figurations of westernness, patriarchy, coloniality—and the illusion of being able to detach from the world that undergirds them—that may accompany it in unquestioned form), it also offers some interesting avenues for epistemological enquiry for approaches that question the loci and modalities of reason. For instance, the Bayesian formulation of subjective prior and posterior distributions might be said to challenge the traditional subject/object dichotomy of Western epistemologies. Equally, a subjective Bayesian approach could be opposed to universal knowledge claims because evidence as measured by Bayes factors is always context-specific and relative.[7]

Subjective Bayesian approaches find resonances with what Marisol de la Cadena has called 'onto-epistemic openings'—analysing modern objects of reason to think what is beyond them but through and with them—or scholars developing feminist science and technology studies have called 'situated knowledge'.[8] Situated knowledge perspectives highlight the embodied and embedded nature of all knowledge claims, where embodiedness is not limited to human embodiment, and embeddedness extends beyond sociopolitical context to include disciplinary perspectives, technoscientific affordances and other more-than-human agencies. Such a position critiques the idea of universal reason and the 'view from nowhere', which too often simply reproduces the false universalism of the white, western, eurocentric, able-bodied, cis-male humanist subject and suggests instead that ethics, epistemology and ontology are always already entangled.[9] Indeed, these categories, whilst crucial as openings to attention, can also themselves become formulaic or over-institutionalised, deferring attention and inhibiting thought whilst nominally addressing problems that are themselves grindingly obedient to such formula. Furthermore, situated knowledges open a more generative possibility of responsibility as 'response-ability', i.e. a capacity to respond and be accountable for the knowledge claims made, and re-figure objectivity as 'positioned rationality'.[10] This is something that resonates with aspects of the embodied, ecological, and enactive approaches to cognition and post-Bayesian approaches to it.[11]

---

[7] Richard D. Morey, Jan-Willem Romeijn and Jeffrey N. Rouder, "The Philosophy of Bayes Factors and the Quantification of Statistical Evidence", *Journal of Mathematical Psychology* 72 (2016): 6–18, https://doi.org/10.1016/j.jmp.2015.11.001.

[8] Marisol de la Cadena, *Earth Beings: Ecologies of Practice Across Andean Worlds* (Duke University Press, 2015). Donna Haraway, "Situated Knowledges: The Science Question in Feminism and the Privilege of Partial Perspective", *Feminist Studies* 14, no. 3 (1988): 575-599, https://doi.org/10.2307/3178066. Sandra Harding, *The Science Question in Feminism* (Cornell University Press, 1986).

[9] Karen Barad, *Meeting the Universe Halfway: Quantum Physics and the Entanglement of Matter and Meaning* (Duke University Press, 2007).

[10] Haraway, "Situated Knowledges".

[11] Andy Clark, *Surfing Uncertainty: Prediction, Action, and the Embodied Mind* (Oxford University Press, 2016).

Situatedness, which generates multiple grounded accountable perspectives, has also been called for by decolonial and postcolonial thinkers arguing for pluriversality[12] and relational empiricism[13] as onto-epistemological and methodological stances or as a philosophical poetics[14]. Foregrounding ontological and epistemological difference, pluriversality is introduced as a decolonial way of addressing the multiplicity of knowledge claims and their geopolitical situatedness as an antidote to western universalism and coloniality of knowledge. Instead, phenomena are understood as relational, materialising through material-discursive practices where multiple agencies might play a role. Arturo Escobar's work on design is particularly relevant here.[15] Design, like mathematics and computing, is a specifically constructed set of ways of grappling with the materialities and processes of the world. Escobar emphasises a multiplicity of dispositions that break up the idea of the possibility of a uniform or universal approach to design. Equally emphasising making, decolonial movements, such as the Zapatistas, look for ways of directly reworking reality at the level of politics, economics and the shaping of relations amongst humans, nonhumans and forms of knowledge, 'from below and for below' as the Sixth Declaration from the Lacondan Jungle puts it, in an exemplary case of thinking in a situated manner.[16]

Given such examples, it is also essential for us to acknowledge the limits and otherwise positioned nature of the argument we are making here. The special issue aims at exploring productive resonances (and dissonances) rather than snug correspondences between situated, pluriversal epistemologies and Bayesian approaches. This might include not only more general convergences and divergences, but also the role of materiality, subjectivity, situated reason, perspective and perspectivalism,[17] measurement, probability, and other relevant concepts. In terms of the context of software studies we are particularly interested in the tensions involved: for instance, in the way in which software may incorporate Bayesian approaches in different ways and at the same time generate a seeming universalism; or the way in which software might incorporate the possibilities for reflexive forms of knowledge and also bring these into productive relation with multiple forms of reasoning or imaginary. There is thus a double movement here, of exploring a sympathy or resonance across the various approaches drawn together, but also reckoning with their differentiation to elicit more about each. We will thus sometimes be presenting a proposition, and then setting out its difficulty. Eduardo Viveiros de Castro, in his explorations of the perspectivism of Amerindian cultures, calls such an approach of working between translations and their deformations, a 'controlled equivocation', comparing translations between fields to walking as a controlled form of falling.[18] These themes run across several articles in this issue.

---

[12] Walter Mignolo, *Local Histories/Global Designs: Coloniality, Subaltern Knowledges, and Border Thinking* (Princeton University Press, 2000). Arturo Escobar, *Designs for the Pluriverse: Radical Interdependence, Autonomy, and the Making of Worlds* (Duke University Press, 2018).

[13] Helen Verran, *Science and an African Logic* (University of Chicago Press, 2001).

[14] Édouard Glissant, *Poetics of Relation*, trans. Betsy Wing (University of Michigan Press, 1997).

[15] Escobar, *Designs for the Pluriverse*.

[16] EZLN http://enlacezapatista.ezln.org.mx/ EZLN, Sixth Declaration form the Lacondan Jungle, June 28, 2005, https://schoolsforchiapas.org › library › sixth-declaration-lacandona-jungle   The question of forming modes of direct action with abstractions and in the abstraction layers of different societies and their computational formations remains an open ground of experiment.

[17] For discussions of perspectivalism see: Friedrich Nietzsche, *On the Genealogy of Morality*, trans. Carol Diethe, ed. Keith Ansell-Pearson (Cambridge University Press, 2017).

[18] Eduardo Viveiros de Castro, *Cannibal Metaphysics*, trans. Peter Skafish (Univocal Press, 2014).

**Schemata of thought**

According to the sociologists Scott Lynch and Bryce Bartlett, the process of Bayesian inference (Eq. 1) 'arguably mirrors the way humans learn about the world: we construct priors regarding how things work and we continually update them with new information as we encounter it.'[19] Yet on the face of it, bringing together Bayes Theorem and situated or pluriversal knowledge might be seen to be somewhat refractory. One is a formula that is more or less precise in its terms. The other two offer idioms for the mappings of precision and inability, of the analysis of power, on the one hand, and capaciousness and difficulty, of multitudes and of openings up, on the other. They are associated with the understanding of knowledge as constitutively political and for the linking of the immediate sensory world with the metaphysical in different ways. This is an odd trio then to bring together. One way to mediate between them would be to employ a schemata, a more or less rudimentary conceptual or procedural mapping, something we do below in the diagram in Figure 3.

Understanding Bayes's method, and subsequent Bayesian techniques, as schemata of thought is also one way in which it can be said to operate. At the point of its introduction in the eighteenth century by statistician and Presbyterian minister Thomas Bayes, the theorem attempts to set out a procedure that would both articulate in general terms a crucial part of the work that is done in the (natural) sciences and offers a means to clarify, and thus make more rigorous what it models.[20] It implies a requirement to stop and think, to provide another register to test thought against other than hunches, training, enculturation, which might otherwise be the primary aspects of directed thought or experiment. It is non-situated, in the sense that it provides an abstract method that may move across different contexts. Its systematisation is what allows it to travel and removes it from too many commitments to a specific configuration of the world or of a problem. As such, it is aimed at being 'non-interested' (to use Isabelle Stengers' term)[21]. By a convolution, this abstraction itself can be said to be what establishes its own particular kind of situatedness. The remove, the abstraction, is what typifies it, makes it of a type, gives it coordinates against which things can be measured. The method as a schemata guides investigation but does not subsume reality. Rather, it may encourage a certain delicacy or carefulness in the making of claims or descriptions. Obversely, the introduction of such a schemata implies an engagement with a material with sharp edges, even numerically edged ones. As a cutting tool, it implies a testing of what is commensurable or incommensurable within any description of a reality that it faces and attempts to describe. This movement between delicacy and the capacity or the aptitude for cutting will mark some of the terrain of discussion here.

As a technique of prediction or (un-)certainty of judgement—and of the modification of prediction or judgement—in relation to the unfolding of an event, what Bayes does is both highly demarcated and compellingly promiscuous. One of the things it cannot do is to fully surmount the question of contingency. Indeed, philosophical writers such as Elie Eyache or Quentin Meillasoux always remind us of the fundamental or foundational force of contingency,[22] especially in those situations that are seen as being nicely described and well under control. The schemata may always be disrupted by an arrival from what it is constrained not to recognise or to encompass. In a sense, situated and pluriversal understandings of knowledge are such strange arrivals. Not exactly the meteorites landing in the mathematician's lap as they press the button to run the program that applies the theorem to the data unfolding from the world through the carefully placed sensors: rather, they might be a sudden recognition of the relevance of entities and factors that are usually excluded from the mixture of things in order to produce the ideal variables with which the experiment is concerned. They bring other kinds of cuts and suppleness to the question. And that is what we hope to achieve with this special issue.

---

[19] Lynch and Bartlett, "Bayesian Statistics in Sociology", p. 50.
[20] Thomas Bayes, "An Essay towards Solving a Problem in the Doctrine of Chances. By the Late Rev. Mr. Bayes, F.R.S. Communicated by Mr. Price, in a Letter to John Canton, A.M.F.R.S", *Philosophical Transactions of the Royal Society of London* 53 (1763): 370–418.
[21] Isabelle Stengers, "Introductory Notes on an Ecology of Practices", *Cultural Studies Review* 11, no. 1 (2005): 183-196.
[22] Elie Eyache, *The Blank Swan: The End of Probability* (John Wiley & Sons, 2010). Quentin Meillasoux, *After Finitude: An Essay on the Necessity of Contingency*, trans. Ray Brassier (Continuum, 2008).

**Outline**

To this end, we first present a brief overview of lines of enquiry into the way mathematics is formed and brings with it the traces of this formation, including those of the societies and medial forms with which it is structurally coupled. We then, in part three, look into the question of functions in machine learning and in Bayes as one way of understanding both the problems being addressed and a means of refiguring them. In part four, we provide a summary of some tendencies in feminist and pluriversal theories that may provide ways of rethinking these. In part five, we provide a summary of the articles for this special issue, and in part six we move on to proposing a few lines for further research.

# 2 Mathematical structures as cultural/social constructs and society as a mathematical form

Mathematics is both a cultural and social construct and a site of invention in which what is produced in mathematics comes to play a part in shaping the possibilities for other kinds of formation, including the social. It is socially constructed in that it is invented and gains significance and possibilities for development in a process of co-composition with wider processes that may call for particular phenomena to become enumerable and calculable. It is inventive firstly in that its modes of thought and action may also be generated by conventions that are more or less loose from direct formation and secondly that what it invents may circulate in multiple forms in the world. This movement between construction and invention is in contrast to the perhaps predominant Platonic view of mathematics as an encounter with divinely or naturally pre-given structures that are simply explored by humans.[23] That mathematics has this double involvement makes it a particularly powerful site for the gestation of operative abstractions and thus also for their potential reinvention or grounding in different modes of reason. The sites of critical intervention here are thus multiple, grounded in the recognition of social and economic forms that encourage or recognise certain kinds of epistemic inventiveness and not others, or developing new social formations that are more politically adequate, and are capable of fighting for, mobilising and inventing new kinds of abstraction.

    The recognition that data is a cultural/social construct has gained traction in recent years, because datafication is a prerequisite for machine learning, and the diversity and garrulousness of its applications has often made their misplaced concreteness clear in its dissonance but appealing in its capacity to produce effects when integrated into wider

---

[23] See: Godfrey H. Hardy, *A Mathematician's Apology* (Cambridge University Press, 2013).

assemblages.[24] This argument has two elements. The first is epistemic: data can never be *raw* as it is not a neutral representation of the world but a construction about and of the world.[25] What is taken to be significant, in what way and by which means shape what comes to be taken as data. The second is political: the gathering of data takes place in particular formations that shape and entrain what is codified. For instance, the concept of 'data colonialism' proposed by Nick Couldry and Ulises Mejias describes a reciprocal dependency between social relations and 'data relations', where the latter are imposed on society by, for example, social media platforms optimising their interfaces for data extraction, structures that in turn may partially replace or improve on, efface or ruin pre-existing or possible social relations.[26] Both construction and invention are operative here.

The mathematics used to produce and analyse data are also deserving of critical attention. We need to ask about the origins and operations of mathematical structures used today, not to discard them immediately but to examine their epistemological commitments and consequences thoroughly. The case for such an approach is strongly made by Alfred North Whitehead.

> *'Nothing is more impressive than the fact that as mathematics withdrew increasingly into the upper regions of ever greater extremes of abstract thought, it returned back to earth with a corresponding growth of importance for the analysis of concrete fact. [...] The paradox is now fully established that the utmost abstractions are the true weapons with which to control our thought of concrete fact.'* [27]

We can add something from the present moment to this observation. Not only does mathematics have significant influence over means of access to the formation of statements or knowledge that are taken to be concrete fact, through computation, it also, more substantially than in Whitehead's day, forms the context in which things come into being as phenomena, whether they are to be pronounced as facts or not.

This motivates us to analyse the mathematical structures at the heart of key elements of machine learning. As part of this process, it is useful to disaggregate the technologies concerned, so that they do not seem as one inevitable and pre-given entity. This approach has the advantage that it is able to recognise the inventiveness that goes into such work, and to see machine learning as an ensemble of techniques which are selected from, modified, and added to through a further process of assemblage.[28] The disadvantage here is that this approach is not quite able to deploy the sweeping scorn of mathematics that some branches of critical theory so enjoy in fully conflating it with a host of the modern banes with which it co-emerged and partially facilitated. Despite this, however, an engagement with the pragmatics and detail of such techniques does not come with the prerequisite of being dutiful to their inception or of doing ordering work on society, work that disavows its conflicts and breaks. Instead, we propose that the struggle for invention, and the invention of abstractions, is a crucial site in which concrete fact, as Whitehead puts it, can be brought into being.[29] At a period in time when such abstractions are often being melded with those described by Alfred Sohn-Rethel as 'real abstractions', in the structures and imperatives of financial capitalism, it may be necessary to work for such inventions outside of disciplinary demarcations and in combination with social forces and imaginaries that have their own logics.[30]

---

[24] Alfred North Whitehead, *Science and the Modern World: Lowell Lectures, 1925* (The Free Press, 1997). Lisa Gitelman, ed., *"Raw Data" is an Oxymoron* (MIT Press, 2013).

[25] Gitelman, *"Raw Data" is an Oxymoron*. Rob Kitchin, *The Data Revolution: Big Data, Open Data, Data Infrastructures & Their Consequences* (SAGE Publications, 2014). Safiya Umoja Noble, *Algorithms of Oppression: How Search Engines Reinforce Racism* (NYU Press, 2018). Ruha Benjamin, *Race After Technology: Abolitionist Tools for the New Jim Code* (Polity, 2019). Yanni Loukissas, *All Data are Local: Thinking Critically in a Data-Driven Society* (MIT Press, 2019).

[26] Nick Couldry and Ulises Ali Mejias, "Data Colonialism: Rethinking Big Data's Relation to the Contemporary Subject", *Television & New Media* 20, no. 4 (2019): 336–49, https://doi.org/10.1177/1527476418796632.

[27] Whitehead, *Science and the Modern World*, p. 34.

[28] Adrian Mackenzie, *Machine Learners: Archaeology of a Data Practice* (The MIT Press, 2017).

**Mathematics as Media**

There is something particular about the nature of mathematical ideas as a form of culture, or as mediatic entities, in that they are passed through history in ways that are often, though not always, relatively unscrambled by the passage of time and the whispering of an idea from one person to another. Philosophical or theoretical concepts, for instance, as this article will no doubt demonstrate, are quite readily slightly misheard, adapted, loosened up or varied as they travel.[31] Mathematical ideas, perhaps because of their frequent parsimony and the requirement for thorough inculcation in certain rites of knowledge before they become intelligible, tend to mutate less readily as they move. One might say that this is due to the concern with a form of precision, one that, as critics such as Whitehead in the quote above have noted, risks loosening them from a relation to the empirical without—given their embedding in particular situations with specific relations of power—a loss to their power, deserved or not, of being determining of their field of application. But a semiotics of mathematics would also look at the different ways in which the notation used to express the idea can change more easily than the idea, according to the conventions of different fields, techniques or periods.[32] A notation becomes a convention at some point, (e.g., the symbols i, j, k usually refer to integers, and capitalized symbols such as A, M–to matrices), but one can express the same idea with different scribbles or with the expressive capacities of different mathematical domains, such as geometry or topology.

This ease of transmission and circulation is assisted by means of the ease of algebraic reference in which one unit of proposition or formula can be referred to or called upon by another proposition, making complexes of mathematical operations readily possible. Indeed, part of the growth of mathematics has been the development of increasingly complex means for integrating kinds of mathematical operations. For instance, at a simple level a formula can 'call' to a function or to the results of one. At another level, a formula might embody more complex systems of reference, such as meromorphic functions that encompass or describe all the possible results of a particular mathematical function. More broadly still, the integration or synthesis of fields of mathematics and the development of their means of reference to each other have been a key aspect of contemporary mathematics.[33] Each mathematical entity becomes a potential building block for others, but also forms sites of attachment in which novel kinds of relation can be formulated through redefinition, subsumption, interfacing with other kinds of question and so on. There is actually a lot of mutation in mathematics, but maybe a mutation just leads to a new construct that joins the population of mathematical constructs, rather than replacing one that is 'less fit'? Such work may also be intercepted by other forms of abstraction, those of capital for instance, which may seek to enablingly direct the probe heads of mathematics, or of the social phantasmagoria that call upon abstractions to willingly or unknowingly go to work.[34] The particular form of mathematics as media yield these possibilities in an ongoing way and crucially means that mathematical culture is not always simply reductive or reifying.

---

[29] Didier Debaise, *Nature as Event: The Lure of the Possible*, trans. Michael Halewood (Duke University Press, 2017). For an important discussion of the power of abstractions, see Mijke van der Drift and Nat Raha, *Trans-Femme Futures* (Pluto Press, 2024). For a discussion of invention, see Gilbert Simondon, *Imagination and Invention*, trans. Joe Hughes & Christophe Wall-Romana (Minnesota University Press, 2023).

[30] Alfred Sohn-Rethel, *Intellectual and Manual Labour: A Critique of Epistemology*, trans. Martin Sohn-Rethel (Haymarket Books, 2021).

[31] Mieke Bal, *Travelling Concepts in The Humanities: A Rough Guide* (University of Toronto Press, 2002).
Michel Serres, *Hermes 1: Communication*, trans. Louise Burchill (University of Minnesota Press, 2023).

[32] Brian Rotman, *Signifying Nothing: The Semiotics of Zero* (Stanford University Press, 1993).

It is partly also in this sense that we are interested in Bayes theorem and its uptake in contemporary approaches to data. A terse but powerful formulation, the theorem has survived its passage from the early eighteenth century in fine working condition, though having been partially buried until the statisticians of the twentieth century found a use for it via a more substantial formalisation of the same principle in the independently achieved work of Laplace.[35] The technique in some ways turns the modern development of hypothesis and experiment based scientific practice into an algorithm. Whilst stated above, the formula for Bayesian inference (Eq. 1) can also be set out as:

*Current Beliefs + New Data = Revised Beliefs*

Such a formula presupposes the supply of data and of actions to vary that data. At the same time, it entails recursion, or a process of feedback between one state and the next that makes the theorem amenable to contemporary imperatives for self-modifying and broadly cybernetic systems.[36] We will show in the next section how Bayes theorem can be used in machine learning (ML) and what mode of reasoning it supports.

# 3 Universal or pluriversal functions as building blocks in machine learning

## 3. 1 The universal function-fitting paradigm in machine learning

What are the mathematical structures at the heart of ML models? Of course, all models require quantification of observed phenomena and so one essential building block of ML is *numbers*, which can correspond to various measurements, counts of occurrences or discrete categories among others. For example, we might be interested in finding the most important words in this special issue and, to quantify this, we could assign each word its total count X. In ML, X is called a *feature* of the observed entities. The task in supervised ML is to predict another quantity Y, called the *target*, from the feature X.

---

[33] Fernando Zalamea, *Synthetic Philosophy of Contemporary Mathematics* (Urbanomic, 2014).

[34] Jacques Derrida, *Specters of Marx: The State of the Debt, the Work of Mourning, & the New International*, trans. Peggy Kamuf (Routledge, 1994). Sohn-Rethel, *Intellectual and Manual Labour.*

[35] See, for instance, Sharon Bertsch McGrayne, *The Theory That Would Not Die* (Yale University Press, 2011). D. R. Bellhouse, "The Reverend Thomas Bayes, FRS: A Biography to Celebrate the Tercentenary of His Birth", *Statistical Science* 19, no. 1 (2004): 3–43.

[36] Yuk Hui, *Recursivity and Contingency* (Rowman and Littlefield International, 2019).

To illustrate the mathematical principles discussed in this section, we propose a toy example, where we are interested in predicting for each word in this special issue the number of articles Y in which it appears from the total count X of the word over all articles. While we do know the total word count and article count of each individual word in this special issue, the task posed here is to learn a potential relationship between the feature X and the target Y that generalises across the observed words. We emphasise that our toy example is not chosen for the scientific value of this task but because it allows us to illustrate the differences between Bayesian and non-Bayesian ML approaches on a dataset where the relationship between X and Y is not known *a priori*.

According to the popular *Elements of Statistical Learning* textbook, *learning* in the context of ML then means to iteratively approximate a *function* that describes the relationship between features X and target Y.[37] This so-called *function-fitting paradigm* puts the spotlight on another important mathematical building block of ML: functions. The model assumption underlying the function-fitting paradigm is that the relationship between target and features is given by a functional relationship

$$Y = f(X) + \varepsilon. \qquad (2)$$

Here the function f maps each possible feature value X (the input) to a target value f(X) (the output) and Eq. (2) states that f(X) recovers the true target Y up to a small (measurement) error $\varepsilon$. Going back to our toy example, this would mean that there exists a function f that determines the relationship between the total count X of a word in this special issue and the number of articles Y it appears in. A problem in ML is that we do not have access to this function f directly, we only assume that it exists. To rephrase the function-fitting paradigm, the goal of ML then 'is to find a useful approximation f̂(X) to the function f(X) that underlies the predictive relationship between the inputs and outputs.'[38] Approximating here means that one tries to get as close as possible to the true function f, even though f is not known *a priori*. The notation f̂ for the approximated function symbolises that f̂ is close to f but might not be the same, hence the symbol ˆ is added to the notation. Given *training data*, which is a set of M observations for both features and targets denoted by $(X_1,Y_1), (X_2,Y_2), \ldots, (X_M,Y_M)$ where the subscripts correspond to the different observations, the quality of the approximation f̂(X) can be measured by computing the difference between the predicted outputs and the true targets, f̂(X) - Y, also called residuals. While it is usually not possible to find an approximation f̂ such that all residuals vanish, one can minimise the so-called *least-squares error* given by

$$|Y_1 - \hat{f}(X_1)|^2 + |Y_2 - \hat{f}(X_2)|^2 + \ldots + |Y_M - \hat{f}(X_M)|^2, \qquad (3)$$

which measures the sum of the squared residuals for all observations.

During the training or learning process, ML models sieve through the training data to 'find the function from the set of admissible functions that minimises the probability of error.'[39] The main difference between supervised ML models is that the 'set of admissible functions' can vary. For example, in the simple linear regression model, the approximation f̂ needs to be a linear function. In artificial neural network models, f̂ can be a composition of many linear and nonlinear functions. It has been shown (in mathematical *universal approximation theorems*) that complex models, such as artificial neural networks, can approximate any continuous function f, earning them the name *universal approximators*.[40] This suggests that ML is an intelligence that can discover all functions—one of the underlying reasons for the current interest in such systems. Before investigating the philosophical consequences of the function-fitting paradigm further, we want to gain more intuition with our toy model.

---

[37] Trevor Hastie, Robert Tibshirani, and Jerome Friedman, *The Elements of Statistical Learning* (Springer, 2009), p. 29.
[38] Ibid., p. 28. The symbol f̂ is read as 'f-hat'.
[39] Vladimir Vapnik and Rauf Izmailov, "Complete Statistical Theory of Learning: Learning Using Statistical Invariants", *Proceedings of the Ninth Symposium on Conformal and Probabilistic Prediction and Applications (Conformal and Probabilistic Prediction and Applications, PMLR)* (2020): 4–40, https://proceedings.mlr.press/v128/vapnik20a.html, p. 31.

As a predecessor of more complex ML models, it is useful to look at linear regression in more detail.[41] The model assumption in linear regression is that f is a linear function of the form

$$f(X) = a * X + b, \quad (4)$$

where the two parameters a and b determine the shape of f. In particular, a determines the *slope* of the linear function f and the parameter b its *constant* or *intercept*. For many contemporary applications, a model may have many such parameters. For our toy example, let us consider as a training dataset the 10 most common words (excluding 'stopwords'[42]) in the 9 other articles within this special issue for which we have obtained total word counts and article counts (numbers based on intermediate versions of articles),[43] see Figure 1. In particular, the word 'machine' corresponds to data-point ($X_1$=132, $Y_1$=7), 'people' to ($X_2$=139, $Y_2$=6), …, and the word 'probability' to ($X_{10}$=331, $Y_{10}$=8). To minimise the least-squares error (Eq. 3), one can obtain optimal parameters â and b̂ analytically, i.e., through a simple mathematical formula.[44] We find that the slope is given by â = 0.016 and the intercept by b̂ = 4.206 and so the least-squares solution of linear regression, f̂(X) = â * X + b̂, tells us that the number of articles in which a (popular) word appears is given by 0.016 times the number of occurrences of the words in the whole special issue plus a constant of 4.206. In Figure 1, the linear regression function f̂ is visualised by a single red line whose positive slope indicates that words that have a higher total word count tend to appear in more articles, as one might expect. How the red curve in Figure 1 seems to cut through the training data, however, seems to emit a kind of authority about the distribution of words in the corpus even when we know that there was no underlying function f that forced the authors to use words according to a certain functional relationship. This may, for instance, be an example of a discursive 'situatedness' in which the vocabulary for the discussion of certain kinds of arguments tends in a shared direction.

---

[40] Kurt Hornik, Maxwell Stinchcombe, and Halbert White, "Multilayer Feedforward Networks Are Universal Approximators", *Neural Networks* 2, no. 5 (1989): 359–66, https://doi.org/10.1016/0893-6080(89)90020-8.

[41] For a critical reflection on linear regression, see, Wendy Hui Kyong Chun, *Discriminating Data: Correlation, Neighborhoods, and the New Politics of Recognition* (MIT Press, 2024).

[42] https://ir.dcs.gla.ac.uk/resources/linguistic_utils/stop_words

[43] A particularity of working with the text of articles whilst they are in the process of peer review is that the articles change, so choosing the moment to carry out this calculation affects the calculation. Additionally, once the text is marked-up for the Word-Press instance used by Computational Culture, and the further mark-up introduced by this software, further changes to what the text of the articles might be taken to be are made. See for further considerations along these lines, Dennis Tenen, *Plain Text: The Poetics of Computation* (Stanford University Press, 2017).

[44] See Wikipedia for more details on linear regression: https://en.wikipedia.org/wiki/Linear_least_squares

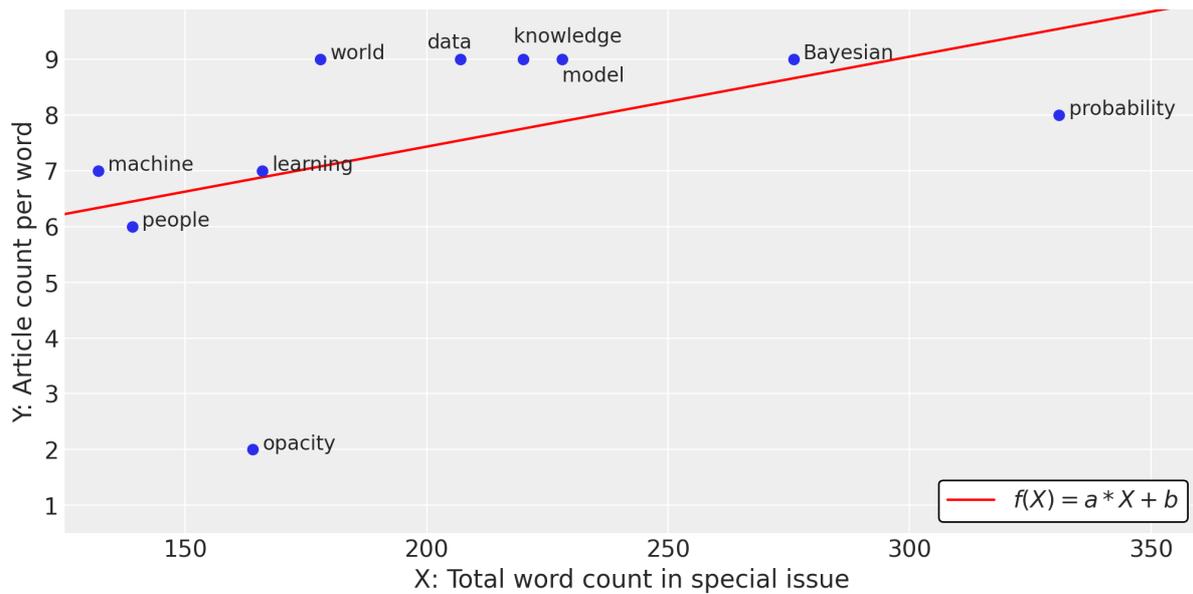

**Figure 1:** *Linear regression applied to the 10 most frequent words in the special issue. As our feature X, we take the total word count in the special issue, and the target Y is the article count per word. We find that the linear function f̂(X) = 0.018 \* X + 3.979 best describes the relationship between the feature and the target (red line). Looking at the diagram, the red line seems to suggest that 'objectively' there is a 'true' distribution of words in this special issue to which all authors adhere. We are interested in undermining the authority of the single line.*

Our illustrations show that the use of mathematical functions in ML is diverse: functions appear in the form of model functions such as linear regression (Eq. 4), loss functions such as the least-squares error (Eq. 3) that quantify the error of the model or gradients, i.e., derivatives of functions that are used for optimization in more complex models such artificial neural networks. According to Adrian Mackenzie, 'functions in machine learning are 'intersocial' in the sense that they bring together different mathematical, algorithmic, operational and observational processes.'[45] While the table (or spreadsheet) has long been used to organise and divide knowledge into fixed categories,[46] functions have the capacity to organise or map more mobile and continuous knowledge. Not only do functions bring together different ML processes (quantification, model formulation, optimisation etc.), but the function-fitting paradigm (Eq. 2) also seems to make an onto-epistemological claim: it implies that there exists a universal function f describing the underlying patterns between features and targets in the first place, and that it can be 'objectively' recovered by f̂ through approximation algorithms. (The degree of objectivity being a slippery quality to determine.) The authority that comes with the approximated function f̂, (e.g., illustrated by the single red line in Figure 1), is thus borrowed from this onto-epistemological claim and leads to overconfident predictions. Before we argue in Section 4 that the function-fitting paradigm is representative of hegemonic Western practices of science that negate other forms of (pluriversal) knowledge, we will first investigate how Bayesian statistics may break open the authority of a single universal function by 'estimating uncertainty' with infinitely many different possible functions as descriptors for the relationship between features and target.

---

[45] Mackenzie, *Machine Learners*, p. 101.
[46] Michel Foucault, *The Order of Things* (Taylor & Francis, 2012), http://public.eblib.com/choice/publicfullrecord.aspx?p=240649.

## 3.2 Bayesians believe in pluriversal functions?

In Bayesian statistics, the parameters of the function f underlying the relationship between features X and target Y are not assumed to be fixed (and just unknown), but the parameters come with inherent uncertainty, which is expressed in terms of (subjective) prior probability distributions. We illustrate this by applying a Bayesian approach to our toy example of predicting article counts Y from total word counts X with linear regression. At first, we can express *Bayesian linear regression* in terms of a likelihood function:

$$Y \sim N(a * X + b, \sigma), \qquad (5)$$

where ~ means that the target Y is distributed according to a *normal distribution* N with mean given by a * X + b, which recovers Eq. (4), and variance σ. The normal distribution, also called *Gaussian distribution*, is a bell-shaped probability distribution determined by its mean and variance.[47] However, in the Bayesian approach, a and b are not fixed parameters but random variables for which we can specify our prior belief. As we have no strong prior knowledge about the slope a, we assume it follows a standard normal distribution, i.e., a ~ N(0,1). For the intercept b, we at least know that it should be positive because a word that appears in the special issue also appears in at least one article, and so we assume that b follows a half-normal distribution, i.e., b ~ HN(1).[48] Finally, we also give a prior distribution to the variance σ (which plays a similar role to the error ε in Eq. (2)) and assume it also follows a standard normal distribution, i.e., σ ~ N(0,1). Note that each pair of possible parameter values (aˆ,bˆ) leads to a single linear regression line fˆ(X) = aˆ * X + bˆ.

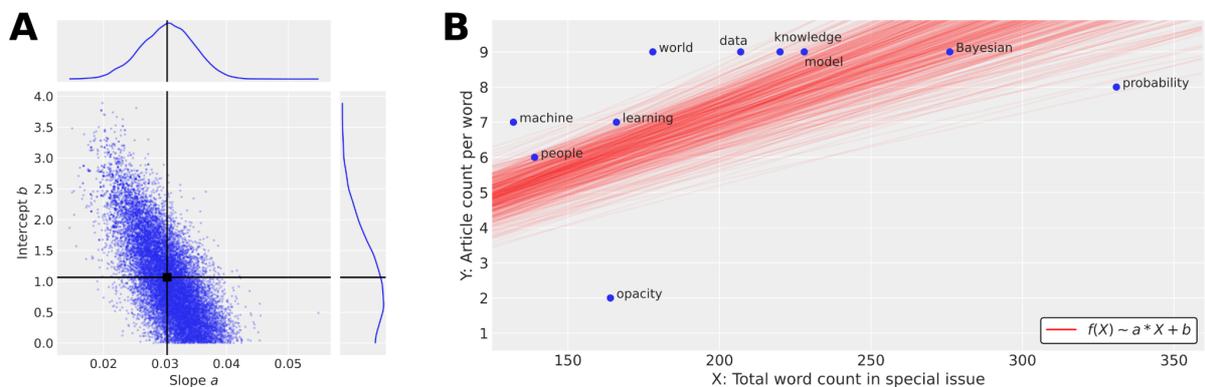

**Figure 2:** *Bayesian linear regression (Eq. 5) applied to our dataset of the 10 most frequent words in the special issue. We use MCMC with the PyMC software to infer the joint posterior distribution p(a,b,σ | Y) of the slope a, intercept b and error σ.* ***A*** *Our MCMC simulations yield 16,000 samples for each parameter, and we scatter plot the samples aˆ and bˆ obtained for the slope and intercept. The spread of the samples illustrates the uncertainty in the parameter values that is captured by our Bayesian model. We can estimate the marginal posterior distributions of a and b (plotted on top and to the right) and find that their mean values (black lines) are 0.030 and 1.065, respectively.* ***B*** *Each sample pair (aˆ,bˆ) corresponds to one possible linear regression line fˆ(X) = aˆ * X + bˆ. Drawing 500 such regression lines (red lines) illustrates again the uncertainty inherent to our inference. While we still observe a generally positive slope (indicating a positive correlation between X and Y), the authority of a single red line (Figure 1) is undermined by the multiplicity of hundreds over other possible lines.*

---

[47] See Wikipedia for more details on the normal distribution:
https://en.wikipedia.org/wiki/Normal_distribution
[48] The half-normal distribution is a fold of the normal distribution at 0 so it only takes positive values, see Wikipedia: https://en.wikipedia.org/wiki/Half-normal_distribution

Based on the counts observed for the 10 most common words $(X_1, Y_1), \ldots, (X_{10}, Y_{10})$, we can update our prior belief in the parameters of the Bayesian linear regression. This means to compute the posterior distributions $P(a \mid Y)$, $P(b \mid Y)$ and $P(\sigma \mid Y)$, i.e., the distribution of parameters a, b and σ conditional on the observations, using Bayes theorem (Eq. 1). Here we use PyMC,[49] a probabilistic programming language (PPL) based on Python for the inference of the posterior distributions with *Markov Chain Monte Carlo* (MCMC) algorithms.[50] MCMC simulations allow us to sample possible values â for the slope, b̂ for the intercept and σ̂ for the variance that follow the joint posterior distribution of a, b and σ. After running MCMC with four independent chains over 4,000 iterations each, we find that the means of the posterior distributions of slope a and intercept b are 0.030 and 1.065, respectively. Note that each sample pair (â,b̂) from the MCMC simulations leads to a different function f̂(X) = â * X + b̂. In Figure 2, we visualise 500 different samples of possible linear regression lines leading to an ensemble of red lines. The ensemble illustrates that there is a lot of uncertainty about the best slope and intercept, although we still observe a generally positive slope (indicating positive correlation between X and Y). Importantly, the authority of a single red line is undermined by hundreds of other possible lines.

Bayesian statistics, thus, has an interesting turn on the function-fitting paradigm. One could argue that the implicit epistemological assumption in Bayesian data analysis is that there is no universal function f describing the relation between input and output in the first place (as posited by the function-fitting paradigm (Eq. 2)) but that there are infinitely many different possible functions f (i.e., relationships) that are all assigned different degrees of (subjective) belief. *Learning* in Bayesian terms then implies the Bayesian updating of your belief in the different possible functions f, using MCMC simulations of the parameters that describe the functions. Hence, a Bayesian linear regression model does not learn a single function but a whole distribution of functions that can be sampled from the 'posterior predictive distribution' as illustrated above (see Figure 2). Breaking the authority of a single line and replacing it with infinitely many alternatives seems to open up possibilities for pluriversal knowledge.

A caveat can be made here that our account so far seems to dubiously imply that pluriversal knowledges can be uncomplicatedly mapped by numerical means. The graph of multiple lines rather than one could, conversely, be an imposition of what John Law calls the One World World (OWW), with the Cartesian coordinate system forming the field of the graph rather than a single line as the integrating framework.[51] Thus, while going from one line to multiple lines might open up some flexibility, the coordinate system itself is still asserted as an integrating system. (David Gauthier's argument in this issue develops a related argument.) The OWW is a version of the world that assigns itself the prerogative to assimilate all others under a protocol of its own devising: a unified field, like the Cartesian grid, which asserts that it contains all options. To make the diagram work, we would have to recognise that it does not work: that there are, for instance, some things that do not readily map to numbers, and that these also make reality. The graph and the procedures it results from would therefore have to be more humble and take its place amongst many other kinds of entity and device, not as a form of relativism, but as a recognition of its mixed potency and impotency. The question then arises, how might mathematical and computational practices arrive at the epistemic humbleness needed, that allows them to take part in the world without tending to dominate it?[52] This is what we intend to discuss next.

---

[49] Oriol Abril-Pla et al., "PyMC: A Modern, and Comprehensive Probabilistic Programming Framework in Python", *PeerJ Computer Science* 9 (2023): e1516, https://doi.org/10.7717/peerj-cs.1516.

[50] We use a version of the Hamiltonian Monte Carlo algorithm, which uses gradients for faster convergence rather than simple random walks. See Matthew D. Hoffman and Andrew Gelman, "The No-U-Turn Sampler: Adaptively Setting Path Lengths in Hamiltonian Monte Carlo", *Journal of Machine Learning Research* 15, no. 47 (2014): 1593–1623.

[51] John Law, "What's Wrong with the One World World?", *Distinktion: Scandinavian Journal of Social Theory* 16, no. 1 (2015): 126–39.

# 4 Feminist and decolonial critiques of mathematical structures

## 4.1 Critical approaches to mathematics

In the last few decades, certain strands in mathematics and especially mathematics education have been undergoing a re-orientation towards decolonisation and social justice.[53] A substantial portion of this shift manifests in acknowledging western biases in the canonical stories of its genealogy, shifting these stories, re-focussing on mathematical traditions in Indigenous cosmologies and cultures as knowledge practices in their own right that need not be 'translated' into western terms for validation and bringing these traditions into pedagogical practice (e.g., through culturally significant algorithms,[54] reworking of curricula and teaching methods[55]). Decolonial theories highlight the colonial structure of power,[56] the operations of which become visible in epistemic operations and forms of organisation through race and gender, among others.[57] Coloniality of power imposes a totalising perspective - what Walter Mignolo calls a 'zero-point epistemology',[58] literally embodied in the measuring tools of imperialism, such as the Greenwich Meridian, from where the British Empire was navigated, or what Haraway calls the 'God trick' of infinite decontextualised, disembodied vision from nowhere/everywhere.[59] Furthermore, both feminist and decolonial perspectives point out that this colonial, eurocentric, phallogocentric structure operates through ontological divisions between what it renders as subject/object, man/woman, nature/culture and others. This has implications for mathematics and its epistemology, in that it decontextualizes and universalizes mathematics, as we pointed out previously.

---

[52] See, Juni Schindler and Matthew Fuller, "Communities as Vague Operators: Epistemological Questions for a Critical Heuristics of Community Detection Algorithms", *Computational Culture* 9 (2023), http://computationalculture.net/community-as-vague-operator/.

[53] J.I. Fúnez-Flores, W. Ataide Pinheiro, A. Ávila Mendoza, R. Phelps and E. Cherry Shive, "The Sociopolitical Turn in Mathematics Education and Decolonial Theory", *London Review of Education* 22, no. 1 (2024): 13, https://doi.org/10.14324/LRE.22.1.13.

[54] Ron Eglash, Audrey Bennett, Casey O'Donnell, Sybillyn Jennings, and Margaret Cintorino, "Culturally Situated Design Tools: Ethnocomputing from Field Site to Classroom", *American Anthropologist* 108, no. 2 (2006): 347–62, http://www.jstor.org/stable/3804796.

[55] e.g., Kate le Roux and Dalene Swanson, "Toward a Reflexive Mathematics Education within Local and Global Relations: Thinking from Critical Scholarship on Mathematics Education within the Sociopolitical, Global Citizenship Education and Decoloniality", *Research in Mathematics Education* 23, no. 3 (2021): 323–37, https://doi.org/10.1080/14794802.2021.1993978. Nandini Bhattacharya et al., "'Being a team of five strong women… we had to make an impression:' The College Math Academy as an intervention into mathematics education", *American Journal of Community Psychology* 70, no. 1-2 (2021): 228-241, https://doi.org/10.1002/ajcp.12573. See also Fúnez-Flores et al., "The Sociopolitical Turn in Mathematics Education and Decolonial Theory".

[56] Aníbal Quijano, "Coloniality and Modernity/Rationality", *Cultural Studies* 21, no. 2-3 (2007): 168-178.

[57] Maria Lugones, "Toward a Decolonial Feminism", *Hypatia* 25, no. 4 (2010): 742-759, https://www.jstor.org/stable/40928654.

[58] Walter Mignolo, *The Darker Side of Western Modernity: Global Futures, Decolonial Options* (Duke University Press, 2011).

[59] Haraway, "Situated Knowledges".

It is not that the totality imagined by such a perspective as a universal is a permanent fixture. Other versions of abusive power, in the mode of the current anti-epistemologies, produce their own devices for establishing meridians, but within the souls of people.[60] These sanctify fears, hatreds, and, above all, resentment as forms of pre-eminent knowing arising from and woven into 'lived experience', a sociological category that has itself become mobile as a form of justification for a reaction trading in the vocabulary of the wounded self. The rigid modes of knowledge classically critiqued by science studies are now supplemented both by the revanchism of those who ache that they lost science's legitimation, and by the post-theoretical strategies of domination employed by those figures of Silicon Valley who warm near to the throne of the new monarchs of ignorance, yet seek to extend the figurations of empire in a more mobile and probabilistic way into the emotions and words of communicants. The old models of reason have been supplemented by, and in some cases broken by, a new kind of delirium of power and vengefulness.

In contrast and in resistance to the coloniality of (western) mathematics, decolonial approaches offer a focus on pluriversality–an approach to ontology and epistemology that highlights multiplicity instead of the aforementioned zero-point epistemology and One-World World. This is expressly explored in Indigenous mathematics and ethnomathematics,[61] a field that studies the formation of mathematical ideas and practices in different cultures, as well as ethnobiomathematics[62] that challenge not only subject/object distinctions, the dualist approach to the understanding of abstract and concrete as mutually exclusive, but also refocus on the entanglements between human and nonhuman, nature and culture in mathematics and mathematics education. Pluriversality also grounds the calls for technodiversity that is intimately connected with pluralization of mathematics.[63] As Baker argues[64], by providing metaphysical grounds for establishing truth and inventing practices for intervening in world-making, mathematics plays an important role in how technosocial environments are shaped. Therefore, it becomes relevant to understand and foster the links between different cultural-civilizational traditions of mathematics and technics, in order to resist technomonocultures.[65]

Decolonial inquiries into mathematics also highlight the role that the discipline played in colonial projects as well as in the Transatlantic slave trade. Practically, through its role in the development of technoscientific advances and applications in fields such as maritime navigation and the operation of plantation economies,[66] mathematics directly contributed to colonial and imperial expansion. Critical race scholars have pointed out how practices of enumeration (in plantation, on the slave ship) positioned black life as an object-commodity, thus intervening not only in epistemic but ontological practices and categorisations.[67] Conceptually, with western mathematics' focus on abstraction, decontextualisation, control and progress, it furthered eurocentric intellectual and cultural hegemony.[68] Increasing mathematisation of all domains of life—i.e. the formatting of life through mathematical means—can also be seen as a kind of imperial/colonial impetus *within* (western) mathematics, which gains even stronger momentum in the age of Big Data and AI as they are formatted by particular forms of capitalism.[69]

---

[60] Matthew Fuller and Eyal Weizman, *Investigative Aesthetics: Conflicts and Commons in the Politics of Truth* (Verso, 2021). Matthew Fuller and Andrew Goffey, *Evil Media* (MIT Press, 2012).

[61] Eric Vanderschiesse and Rik Pinxten (eds.), *Indigenous Knowledge and Ethnomathematics* (Springer, 2022).

[62] Ron Eglash, "Ethno-biomathematics: A Decolonial Approach to Mathematics at the Intersection of Human and Nonhuman Design", in *Ubiratan D'Ambrosio and Mathematics Education*, ed. Marcelo C. Borba and Daniel C. Orey (Springer, 2023), https://doi.org/10.1007/978-3-031-31293-9_18. Ethnobiomathematics is a field that points to how the work of mathematical knowledge is performed not only by humans but by non-humans as well. As Eglash points out, the 'framework of ethno-biomathematics allows us to incorporate Indigenous perspectives in which knowledge is co-produced by human and nonhuman agency' (p. 289).

[63] Yuk Hui, *Machine and Sovereignty: For a Planetary Thinking* (University of Minnesota Press, 2024).

[64] Michael Baker, "The Western mathematic and the ontological turn: Ethnomathematics and cosmotechnics for the pluriverse", in *Indigenous Knowledge and Ethnomathematics*, ed. Eric Vandendriessche and Rik Pinxten, (Springer, 2022), 243–76.

[65] Technics is a term often used in philosophy of technology to include both technology and techniques.

Feminist approaches to mathematics often centre on mathematics education. They highlight how inequalities of gender and race permeate mathematics education, as well as more broadly how power dynamics, identities and structural inequalities are at play in the classroom[70]. Scholars, such as Suzanne Damarin, argue that there is a lack of engagement by feminist theories with the fundamentals of mathematics. Partly this is, of course, because of the presumed universalism and disembodied, decontextualised rationalism: if mathematics and, by extension, rationality are universal, and if mathematics has a direct relationship to truth, then gender (nor other contextual or political factors) should make no difference. Despite this, Damarin explains the gap between feminist theory and mathematics by the observation that mathematics often acts as a 'marker in the lives of women'[71], reproducing structures that entrain feelings of incompetence or guilt. Laura Black gives further evidence how binary gender operationalisation and the coding of mathematics as male leads to oppressive learning environments at school. However, through interviews as part of the 'nonbinary maths project', Black also found that mathematics can act as an empowering 'third space' for some queer, trans and gender non-conforming students.[72]

As an overarching epistemological position, aspects of mathematics have been criticized by feminist philosophers of science, such as Donna Haraway, Sandra Harding,

---

[66] E.g., see Amir Alexander, "The imperialist space of Elizabethan mathematics", *Studies in History and Philosophy of Science* 26, no. 4 (1995): 559-591; S. Khan, S. LaFrance, and H. T. T. Tran, "After plantations' precarities: curating math-thematic curriculum plots in initial teacher education for multispecies' flourishing and a freedom-yet-to-come", *Research in Mathematics Education* 24, no. 2 (2022): 170–186, https://doi.org/10.1080/14794802.2022.2090421.

[67] Katherine McKittrick, "Mathematics Black Life", *The Black Scholar* 44, no. 2 (2014): 16-28.

[68] Alan J. Bishop, "Western mathematics: the secret weapon of cultural imperialism", *Race & Class* 32, no. 2 (1990): 51-65. Diane M. Nelson, *Who Counts?: The Mathematics of Death and Life after Genocide* (Duke University Press, 2015).

[69] Eva Jablonka, "Mathematisation in environments of Big Data – 'implicit mathematics' revisited", in *Proceedings from CIEAEM 69, Chapter 1: Plenaries*, ed. B. di Paola and U. Gellert, *Quaderni di Ricerca in Didattica (Mathematics)* 27, suppl. no. 2 (2017): 43-51, https://sites.unipa.it/grim/quaderno27_suppl_2.htm. Anna Chronaki and Dalene Swanson, "De/mathematising the political: bringing feminist de/post-coloniality to mathematics education", in *Quaderni di Ricerca in Didattica' QRDM (Mathematics)* 27, suppl. no. 2 (2017): 67-71.

[70] Linda McGuire, "Feminist Theories Informing Mathematical Practice", in *Handbook of the Mathematics of the Arts and Sciences, ed.* B. Sriraman (Cham: Springer, 2020), https://doi.org/10.1007/978-3-319-70658-0_77-1.

[71] Suzanne Damarin, "Toward Thinking Feminism and Mathematics Together." *Signs: Journal of Women in Culture and Society* 34, no. 1 (2008): 101–23, https://doi.org/10.1086/588470, pp. 117.

[72] Black et al. "The connection between gender and school mathematics: the views of queer, trans and gender non-conforming students" (2025).

Evelyn Fox-Keller and Helen Longino. In particular, Haraway questioned the idea of universal knowledge and decontextualised scientific objectivity through her concept of 'situated knowledges', proposing that science should embrace perspectivism and reframe objectivity as positioned rationality[73]. Furthermore, Haraway also drew attention to the epistemic role that embodiment and technology play in knowledge endeavours—a position that has been later taken up and expanded on by Karen Barad.[74] She also highlighted the importance of accounting for multiplicity and non-closure of situated locations, including subjects and objects of knowledge. According to Haraway, '[s]ituated knowledges require that the object of knowledge be pictured as an actor and agent, not as a screen or a ground or a resource'.[75] Her work thus constituted an account of feminist objectivity and feminist technoscience that is both locatable and accountable, yet resistant to fixity: feminist technoscience has to do with boundary work because both subjects and objects are in constant relation and flux.

Black feminist theorists, such as Patricia Hill Collins[76] have also highlighted the contextual and social situatedness of knowledge claims and their entanglements with power and intersecting structural inequalities. Concepts, such as 'strong objectivity'—grounding knowledge practices in peoples' lives rather than in an assumed 'neutral mode'—have been introduced by Sandra Harding in her explorations of feminist and postcolonial perspectives in science and technology.[77]

Feminist epistemologies often focus broadly on embodiment and embeddedness, which is to say materiality, situatedness and entanglements, providing grounds for some existing feminist critiques of the epistemologies and practices of mathematics. For instance, Bonnie Shulman traces some of the fundamental axioms in mathematics and their sociocultural histories and argues for a more complexified approach to mathematics that takes into account how 'values are encoded even in the *logical language itself*'.[78] In a more practices-oriented work, Sara Hottinger investigates how mathematical ways of knowing are taking shape within communities, and how mathematical subjectivity gets shaped by various cultural norms around gender and race, among others–and vice versa, showing how mathematics constructs normative western subjectivity and plays a role in the constitution of the west itself.[79]

Another notable intervention into the ontological and epistemological questions in mathematics, situated both in feminist and decolonial fields, is the work of Helen Verran, who investigated how differently situated cosmologies germinate different worlds and thus also different mathematics. Verran's work is exemplary of a kind of *relational empiricism* that proposes that research objects emerge in materialsemiotic relations, not least of which are research methods and assemblages.[80] Resisting both the straightforward argument of the equivalence of different mathematical traditions, which would leave them again reducible to a monolithic universal, and cultural relativism that would position different knowledges as insular and separate, Verran instead navigates the composition and decomposition of relations, the sociocultural and material making of numbers and insists on creating space for epistemic pluralism and epistemic flourishing[81] as an ongoing practice.

---

[73] Haraway, "Situated Knowledges".
[74] Barad, *Meeting the Universe Halfway*.
[75] Haraway, "Situated Knowledges," p. 592.
[76] Patricia Hill Collins, *Black Feminist Thought* (Routledge, 1990).
[77] Sandra Harding, "Rethinking Standpoint Epistemology: What is 'Strong Objectivity?", in *Feminist Theory: A Philosophical Anthology,* ed. Ann E. Cudd and Robin O. Andreasen (Oxford: Blackwell, 2005). Sandra Harding, "Postcolonial and feminist philosophies of science and technology: convergences and dissonances", *Postcolonial Studies* 12, no. 4 (2009): 401-421.
[78] Bonnie Shulman, "What if We Change Our Axioms? A Feminist Inquiry into the Foundations of Mathematics", *Configurations* 4, no. 3 (1996): 449, https://doi.org/10.1353/con.1996.0022.
[79] Sara N. Hottinger, *Inventing the Mathematician: Gender, Race and Our Cultural Understanding of Mathematics* (State University of New York Press, 2016).
[80] Verran, *Science and An African Logic*.

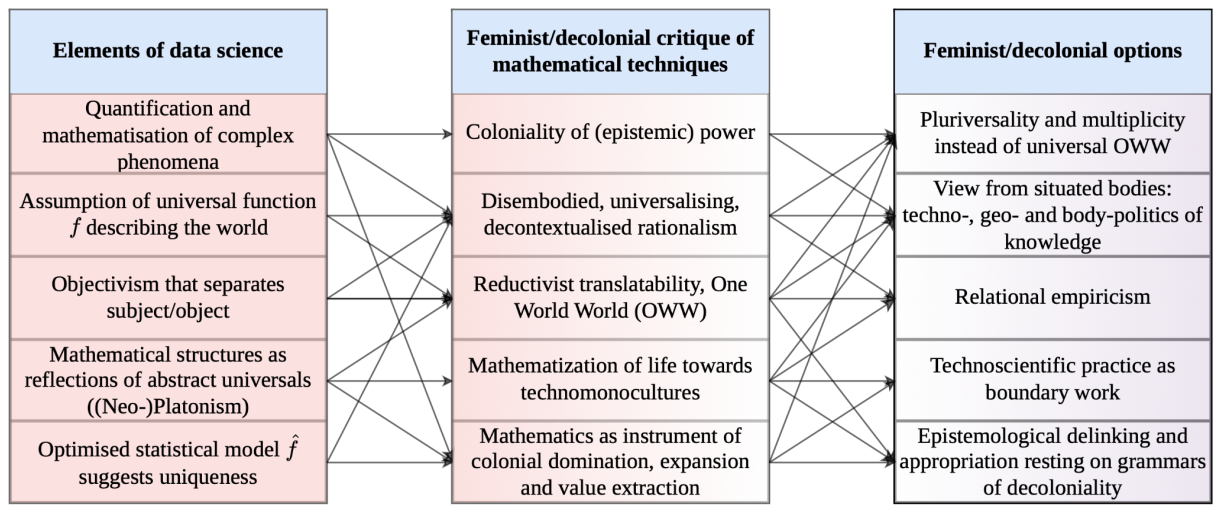

**Figure 3:** *In a diagrammatic form that resembles the architecture of an artificial neural network, the figure summarises different characteristic elements of machine learning in the first (vertical) layer, their feminist and decolonial critique in the second layer and oppositional feminist and decolonial options in response to Western epistemologies in the third layer. The different arrows indicate implications or responses. Of course, the figure is only a very simplified and distorted representation, because it literally bounds arguments from feminist and decolonial theories into separate boxes.*

In Figure 3 we try to indicate, in a few lines, the kinds of things that might figure in a mobilisation of Bayes in relation to the theoretical work we are interested in here. Of course, this is rather the wrong thing to do in this context in that, as Donna Haraway suggests, such work should be about 'heterogenous multiplicities that are simultaneously salient and incapable of squashed into isomorphic slots or cumulative lists'.[82] A tabulation cannot encompass true multiplicity, which is always more than being simply several. Haraway notes that such figures offer an illusory symmetry, implying that the positions offered are alternatives to and exclusive of each other. Lists and tables are nevertheless mobilised in Haraway's work as a way of showing incompleteness and multidimensionality, with the caveat that the dimensions of a thing are rarely simply exhausted by their enumeration. It is in this spirit that we offer Figure 3.[83] The question of the table, working with categories, and the vector, working with the projection of open-ended functions, is discussed at length in Adrian Mackenzie's *Machine Learners*.[84] This book provides a crucial inspiration for thinking with formulae such as Bayes.

## 4.2 Situating Bayes, first steps

In some sense, Bayesian epistemology and its operationalisation through Bayesian data analysis still largely follows the trajectory of Neoplatonic data science. Western epistemic authorities of science have raised many objections against subjective Bayesianism. For instance, subjective Bayesian statistics is often discounted as non-scientific and lacking objectivity (in a conventional meaning of the term). In machine learning therefore, the supposedly more objective 'noninformative priors' (instead of subjective priors) are usually used,[85] while prior distribution might be understood only as a 'regularisation device'[86] that helps keep parameter values in a certain range. Nonetheless, we suggest that there is space to foster more general awareness of the subjective Bayesian framework as potentially germinal of resonances with feminist and decolonial epistemological projects—this special issue is indeed an attempt to open such space.

---

[81] Martha Kenney, "Counting, Accounting and Accountability: Helen Verran's Relational Empiricism", *Social Studies of Science* 45, no. 5 (2015): 749-771.
[82] Haraway, "Situated Knowledges," p. 586.
[83] For an extended discussion of the table given in Haraway's 'Cyborg Manifesto' see, Zach Blas, Melody Jue, and Jennifer Rhee, eds., *Informatics of Domination* (Duke University Press, 2025).
[84] Mackenzie, *Machine Learners*.

A recognition should be made here that this question is not just epistemic but also has, for instance, substantial significance in terms of the environmental impact of some of the technologies discussed in terms of their use of energy, production of greenhouse emissions, and use of resources such as water. These questions are not addressed here, but it is notable that the splitting of knowledge of the world where the subject is bracketed off from what is constituted as its object has significant consequences. Knowledge formations understood as a literal part of the world, whether as ideas, words, or as data centres, take ecological form.

We now thus turn to suggesting some preliminary directions for mapping resonances between Bayesian practices and, broadly speaking, aspects of feminist perspectives, including feminist relational empiricism, the formation of pluriversalities and decolonial approaches. These can be grouped into three aspects: situatedness, relationality and multiplicities. Situatedness pertains to the situated, embedded and embodied notion of knowledge claims, relationality highlights entangled processes of knowledge making, and multiplicities point to knowledge as a practice of boundary-making.

**Situatedness**

On a meta-theoretical level, situatedness and locatability resonate with Bayesian epistemology and can possibly be operationalized through Bayesian practices as a way of *explicitly* taking locality into account and thus making machine learning projects accountable. After all, much of the lack of accountability in technoscience rests on the 'trick'—the aforementioned 'God trick'—of bracketing out location and, through imperatives for technology to scale, removing the need to contextualize. Situatedness, then, would need to go beyond the formal computation of prior and posterior distributions and find a way to account for instability and openness of location, their internal multiplicity and flux.

That is to say that Bayesian formulations of subjective prior and posterior distributions may be one way to help challenge the subject/object dichotomy of western epistemologies, when subjective prior distributions are used that represent or articulate the embodied, or otherwise situated, context or experience of people or other entities with *epistemic privilege* in a certain issue. There is a tension in using these terms in that mapping something in numerical terms carries its own risks of abbreviation and foreshortening. Nevertheless, transposing a situation into another media may give a means to grapple with it, and even the difficulty of the mapping operation may make some things more amenable to understanding.

In these terms, there might be room for acknowledging situated knowledges in modelling and mathematisation. Part of the effect of this grappling with transposition would be in reckoning with the multiple tendencies forming the positionality of the researcher. Such a positionality can be partially represented as a prior distribution. One would have to arrange for numerous contextual factors to be understood and given a numerical weighting. For an experimenter to map their positionality introduces the old philosophical problem 'know thyself'. It is an important process but simply shifts the locus of the question if the prior is taken to be a matter of identity. Rather, it can be an invitation to becoming. Moreover, 'knowledge elicitation' techniques have been developed that help 'experts' (individuals or communities) with the difficult task of encoding their knowledges in terms of prior probabilities.[87] While Anthony O'Hagan et al. follow a very pragmatic approach in the book *Uncertain Judgements: Eliciting Experts' Probabilities*, where, if '[t]he person whose knowledge is to be elicited is usually referred to as an 'expert''[88], one could also reinterpret the concept of an 'expert' through a feminist lense and take the 'expert' as the person with 'epistemic privilege'. This taken into account, prior distributions could make visible aspects of the 'geo- and body-politics of knowledge' as required for epistemic decolonisation according to Mignolo.[89] Some such data is indeed brutally sited in bodies and spaces.

---

[85] Gelman et al., *Bayesian Data Analysis*.
[86] Andrew Gelman and Cosma Rohilla Shalizi, "Philosophy and the Practice of Bayesian Statistics", *British Journal of Mathematical and Statistical Psychology* 66, no. 1 (2013): 8–38, https://doi.org/10.1111/j.2044-8317.2011.02037.x.

**Relationality**

Some of it might have to do with approaching Bayesian practices in their own infrastructural, contextual situatedness: if we take seriously Verran's proposition that objects materialise in relations, then it is important to not lose sight that Bayesian techniques are not objects-as-such but clusters of relations that get configured differently in different sociotechnical settings. Such a relational perspective and mode of practice would also be needed in order to account for the relationality of the kinds of mathematico-social assemblages that Bayesian practices produce.

Another example would be measuring evidence with Bayes factors (the ratio of posterior distributions of two competing statistical models), which are always context-specific and relative.[90] This would entail seeing the mathematics or the software into which it is more or less transcoded (not simply the human person figured as calculator and reasoner) as situated and the interplay of forms, rather than a fixed result being the crucible of the situation. Indeed, a variant of this article could be written that would place this as the central point of emphasis, from the perspectives of calculations.

A further alternative would be to see the operation of a calculation not as one of self-mapping, but of invention, where the problem at hand may partially or substantially transform in ways that are more or less correlated with its mathematisation. To reframe our proposition in these terms would be to argue that systems of symbols produce something like a landscape that can be iteratively explored. This would extend to processes of ideation and reasoning which provide particular terrains that place limits on action and affordances for it, but also inflect and condition it in some ways. Poetic writing provides different ways of inhabiting and producing such a textual landscape, and they are often rich with seams of nonsense and noise. Symbolic systems can be inhabited in a way that produces paradox in a way that has an affinity to mathematical paradoxes. Daniil Kharms's poem about the red-headed man, who exists purely in supposition exemplifies this: the description of the person is built up, by nose, by ear, by mouth, by the text which names and describes them. The poem then evaporates them. It is a diagnosis of the capacity of text, in combination with an imagining reader, to bring into being things that are not there.[91] Something similar can be seen in landscapes of calculation and of code.

---

[87] Anthony O'Hagan et al., *Uncertain Judgements: Eliciting Experts' Probabilities* (Chichester: John Wiley & Sons, Ltd, 2006).
[88] Ibid., p. 9.
[89] Walter Mignolo, "Delinking - The Rhetoric of Modernity, the Logic of Coloniality and the Grammar of de-Coloniality", *Cultural Studies* 21, no. 2–3 (2007): 449–514, https://doi.org/10.1080/09502380601162647.
[90] Morey, Romeijn, and Rouder, "The Philosophy of Bayes Factors and the Quantification of Statistical Evidence".

Such capacities do not need to be made of simply one thing. Félix Guattari describes—taking a phrase from mathematician Henri Poincaré—'abstract machines', as looped together from multiple kinds of materials: media, hormones, ideas, habits, codes, technologies, all existing for a moment or gradually accreting or shifting over the course of a lifetime.[92] Inhabiting and participating in this kind of multiform landscape as a situation characterises many contemporary digital media formations. A key attraction of machine learning and related techniques is their potential ability to map, approximate and recompose aspects of such multidimensional abstract machines, given means for their enumeration. Guattari's formulation is also relevant in that it is fundamentally social or relational, as is the given self or situation. But it also recognises that the social is also folded through persons and other formations, such as media or software, whether configured as subjects or not, and that these have the power, indeed in many ways, the inevitability, of rearranging things as they pass through them. Mathematics as a form of thinking and communicating that can be subtly attuned to transformation can help us experience and to think with this and may in turn provide grounds for other forms of mathematics, for a machine learning that would be more aligned with situated and pluriversal perspectives. The mathematics of Bayesian statistics might open such a way of folding into machine learning models the subjective beliefs or conceptual experiences of those who develop, train or use the model.

**Multiplicities**
A convergence around the issue of multiplicity is another possible resonance. As we have seen from simple examples, such as the Bayesian regression of keywords from this special issue presented earlier, Bayesian techniques can indeed account for some degree—or rather some *type*—of multiplicity. As such, we need to talk about multiplicities in a plural sense.

To start with, the focus of Bayesian approaches on a multiplicity of functions instead of one correct one constitutes the first possible direction. One line into this question is temporal: the distinction between input and output, the prior and the posterior, may work in some cleanly demarcated contexts, but is already an awkward translation of many processes. Methods to measure confounding factors are brought in to provide forms of articulation for this difficulty, but may, according to context and formulation, simply restage it at a finer degree of granularity.

The epistemological assumption of a universal function lands differently according to context. In certain domains like physics, this assumption might be a simplification but perhaps one that is not so problematic. In other domains like social sciences and their interweaving with vernacular lives, however, the idea of a universal function can pave the way to the negation of other forms of knowledge (those not aligned with the authority of a universal function) and can result in what Gayatri Spivak calls epistemic violence on the one hand, and authoritative-sounding formulations with wispy levels of traction on the other.[93] A Bayesian approach with a focus on the predictive posterior distribution from which you can sample a variety of possible functions (see Figure 2B) might, in some contexts, oppose universal knowledge claims and be able to play a part in acknowledging pluriversality.

---

[91] Daniil Kharms, "Blue Notebook no.10", in *Today I wrote Nothing: The Selected Writings of Daniil Kharms*, trans. Matvei Yankelevich (Ardis, 2009), p.45. See online text in English here.
[92] Félix Guattari, *Chaosmosis: An Ethico-Aesthetic Paradigm*, trans. P. Bains and J. Pefanis (Power Publications, 1995).

Moreover, given very diverse prior knowledges, it might be possible to follow a *non-parametric* approach to Bayesian data analysis, which, according to Zoubin Ghahramani 'encompasses an open-ended universe of models that can adapt in complexity to the data'[94] and includes the application of *Gaussian* or *Dirichlet process models*,[95] whose description lies beyond the scope of this article. The open-endedness of non-parametrics could potentially be mobilised for opposing narrow epistemologies and realise its own kind of pluriversality of different knowledges given the recognition of the interplay of translations and mappings inherent to such a project.

Pluriversality and multiplicity[96] in feminist and decolonial epistemologies means being grounded in differential world-making practices as well as being internally non-unitary and open-ended. How does Bayesian enumerative multiplicity compare and contrast with these epistemologies? Would it be possible to argue, for instance, that the features and targets that represent the phenomenon that is measured or analysed (e.g., the X and Y in Eq. 2) render the phenomenon fixed, and what a Bayesian approach then allows is an account of a differential *embedding* of the measured phenomenon in the world–in a sense of how a phenomenon is positioned in the world and how it relates to other elements in its milieu? Might it then mean that the dissonance between Bayesian and feminist technoscientific perspectives would be that the former does not allow for intra-phenomenal changes, whereas the latter would focus on such changes through exploring entanglements? Alternately, one could also argue that Bayesian techniques are more attuned to intra-phenomenal changes, because they allow for iterative updating of the posterior whenever new observations are made, even if the likelihood which defines the model stays the same. The question thus remains somewhat open and perhaps, again, would need to be situated in a specific practice and actualisation of Bayesian epistemology.

## 5 Summary of contributions to special issue

This special issue arises from a workshop on 'Bayesian Knowledge' at the Digital Culture Unit at Goldsmiths, University of London, in November 2023. The workshop included participating researchers with backgrounds in mathematics, philosophy, computer science, media arts and sciences, cultural studies, interdisciplinary design, media theory, sociology, science and technology studies, art, political science, posthumanism, machine learning, software studies, and other fields.[97] In all of these papers we see vivid proposals for rethinking and reworking computational practices, as follows.[98]

---

[93] Gayatri Chakravorty Spivak, "Can the Subaltern Speak?", in *Marxism and the Interpretation of Culture* (London: Macmillan Education, 1988), 271–313.

[94] Zoubin Ghahramani, "Probabilistic Machine Learning and Artificial Intelligence", *Nature* 521, no. 7553 (2015): 452–59. Quote on p. 454.

[95] Gelman et al., *Bayesian Data Analysis*. See also, Carl Edward Rasmussen, "Gaussian Processes in Machine Learning", in *Advanced Lectures on Machine Learning: ML Summer Schools 2003, Canberra, Australia, February 2 - 14, 2003, Tübingen, Germany, August 4 - 16, 2003, Revised Lectures*, ed. Olivier Bousquet, Ulrike von Luxburg, and Gunnar Rätsch (Berlin, Heidelberg: Springer, 2004), 63–71.

[96] Arturo Escobar, *Designs for the Pluriverse: Radical Interdependence, Autonomy, and the Making of Worlds* (Duke University Press, 2017). Verran, *Science and an African Logic*. Haraway, "Situated Knowledges". For an operationalization of intersectional feminist principles of pluriversality and multiplicity in data science, see D'Ignazio, Catherine, and Lauren F. Klein. *Data Feminism*. The MIT Press, 2020.

Clemens Apprich's paper *Learning How to Learn. Abduction as the 'Missing Link' in Machine Learning* investigates abduction as a function of learning and asks whether Bayesian epistemology and Bayesian machine learning techniques could be seen as an abductive method, linking symbolic (deduction based) and connectionist (induction based) AI. Abduction, Apprich argues, is a key process in learning–all learning, including machine learning, is abductive because in some way or another, it has to rely on conceptual knowledge, either directly (in case of human learning) or indirectly through reliance on domain expertise, structural representations that data are imbued with and so on. However, only Bayesian approaches, according to him, explicitly rely on abduction by taking into account the conditions in which specific knowledge is generated, i.e. by accounting for prior beliefs and emerging evidence in probability estimation. Apprich then looks into Bayesian belief networks as an example of such abductive learning, proposing that they 'can serve as a formal mechanism to support and enhance abductive reasoning'. Calling for more explicit attention to and inclusion of the situated and contextualized conditions of machinic learning, Apprich suggests that abduction and its uptake in Bayesian epistemology can provide a framework for intelligent machines to 'learn how to learn' towards more creative and speculative machinic reasoning.

In their paper *Computing a Pluriversal Future*, Alan F. Blackwell, Nicola J. Bidwell and Charlie Nqeisji investigate a pluriversal lens in the domain of probabilistic programming languages (PPLs). PPLs are used by statisticians to express Bayesian models and to use those models for inference, estimation, prediction and other purposes. The authors compare this statistical modelling approach to the ways of making decisions under conditions of uncertainty as practised by the Ju/'hoansi people in the Nyae Nyae conservancy of the Kalahari. The paper brings together tools and observations from different disciplinary backgrounds, namely computational and ethnographic work, as 'an invitation for the reader to take these as pluriversal perspectives on Bayesian knowledge'. As a concrete illustration of contrasting modes, they first present stories around chance and probability told by Ju/'hoansi collaborators and the spinner game objects that facilitated these discussions. Then the authors move to introduce the Multiverse Explorer, a PPL designed to be accessible to non-technical users. The authors argue that PPLs can offer an alternative to the neural network-based AI in that they engender possibilities for transparency, alignment and explanation and thus enable explicit deliberation of underlying policies and explanations. In the last step, the authors juxtapose their experiments with PPLs for Western audiences with their investigation of Bayesian reasoning with Ju/'hoansi collaborators. In relating the Multiverse Explorer prototype to the pluriversal knowledge systems among diverse communities, the authors consider ways in which more accessible Bayesian models might offer an alternative to the established inequities of epistemic injustice.

As a way of broadening the scope of the issue, we also include an interview with Maria Chehonadskih, author of *Alexander Bogdanov and the Politics of Knowledge After the October Revolution*.[99] Chehonadskih's book traces the way in which the politics of knowledge was conceived of and worked on in the late nineteenth and early twentieth century in a Russian and Soviet context. This rich account follows the way in which epistemology, following the theories of physicist Ernst Mach, through the work of philosopher and activist Bogdanov, to the literature of Andrey Platonov, became a lively and inventive field in which the new kinds of subjectivity imagined or produced through the revolution might be conceived.  What kinds of perspectivalism might be created by a comrade-machine or, thinking ecologically, by a comrade-animal?  How does the experience and knowledge of technology, or the formation of a class, change what it is to know?  As the start of the Soviet era shaded to something more deadly, these questions took on a different tone, and Chehonadskih's book is a major attempt at reviving questions buried under the weight of history.  Resonances with contemporary questions about the politics of knowledge in relationship to power and technology are artfully developed by Chehonadskih in this interview.

---

[97] The different disciplines are listed in a random order.
[98] Papers in this special issue are ordered alphabetically according to the name of the first author.
[99] Maria Chehonadskih, *Alexander Bogdanov and the Politics of Knowledge After the October Revolution* (Palgrave Macmillan, 2023).

In the article *Quantum Cognition and the Limits of Classical Probability Models*, Elizabeth de Freitas discusses some of the onto-epistemic problems with classical probability, and explores alternative ways of formalizing modes of plausible reasoning. De Freitas suggests that doubt and dependency are mischaracterized in conventional formalizations of conditional probability, rendering them inadequate in their capturing of contingency by the kind of "procedural randomness" that can be seen in the operations of neural networks. As an alternative, de Freitas discusses quantum probability and the emerging field of quantum social science that addresses alternative quantum-like mathematical formalizations of reasoning under uncertainty. De Freitas shows that quantum probability envisions a multi-dimensional state of potentiality where incompatible quantum futures are in superposition, problematizing the Bayesian reliance on calculating joint probability distributions.

*The Cancelling Out of Chance*, David Gauthier's contribution to this issue, addresses calculations of probability through a set of questions grounded in the philosophy of mathematics and of chance alongside the historical genesis of techniques of probability.  David Hume's sceptical approach to mathematical accounts of the world, elaborated by de Finetti, provides entry points into a consideration of how the fundamental contingency of the universe may be both calculated and elided.  Here, there is a challenge to the question of the subjective and its foundations.  By elaborating the tools it needs to address the world mathematics complexifies and gains its own characteristics and self-consistency that differentiate it sufficiently from the world that it may refer to.  As such, the formation of self-consistent structures of axiomatics differs from epistemology.  Chance or contingency is replaced by a structured relation of rationality against which perception must be calibrated, that is "the assumption that the world has a tendency to repeat itself in the long run"[100]. An inherent movement from mathematics to metaphysics is thus entailed where 'rational beliefs and world(s) seem to be to be structured like the 'language' of probabilities'.  This language in turn, as a matter of course, serves as a mechanism of exclusion.  Barbara Cassin's recovery of the potential of sophistry as a parallel form of philosophy provides a way of understanding this formation of a perfect language, one that never arrives but that, along the way, can still have powerful effects.  In that mathematics can do things, without ever reaching totalisation, an opening is made, indeed is necessitated, towards the question of pluriversality.

Christopher Lawless's article *Subversive Witness: The Disruptive Influence of Bayes Theorem on Forensic Science* addresses the use of Bayesian procedures of reasoning in the evaluation of forensic evidence in legal processes.  The article draws on two examples of criminal assault investigation and DNA evidence to show how, in different ways, situated knowledge acts not as a panacea but as an inevitable coefficient of all knowledge, sometimes involving tensions and entanglements that should prompt ethical and political reflection.  The wider analytical structure and societal forms in and through which the calculation takes place plays a part in determining what is opened up or rendered silent. Here, there is a tension between the formal description of Bayesian methods and their application in specific practices. These may enfold forms of prejudice and the operation of unspoken norms with abusive social consequences. The lacunae of investigations may be formed by intersubjective qualities, woven into technical systems and working processes, rather than simply subjective ones.

---

[100] See David Gauthier, this special issue.

Adrian Mackenzie's article, *Generating Samples: Re-writing Bayes Rule as a Probability Hack* is written with a certain kind of Deleuzian humour. The article proposes understanding Bayesian approaches in a number of ways, firstly by working with them in practical terms through sampling procedures, secondly by thinking through the question of approximation in structurally complex contexts and thirdly by thinking through the questions of approximation, recognising the transitions, translations and deeply cultural interconnections of the different operations they might make in contemporary contexts. In his opening discussion, Mackenzie notes that some of the key gains of DeepMind, as an example, have been in developing new ways of linking systems of images to those of statement production. Finding new means of connecting the seeable and the sayable, and turning the latter into programmatic forms, characterises many of the recent gains in the sciences of the artificial. One such new means is by inhabiting the probability distributions of Bayesian procedures by a hack. The interwoven cascade of probability distributions in machine learning at the industrial scale of large language models seemingly challenges inhabitation by preemptively overwhelming it. Mackenzie proposes reframing this condition by constructing new wagers: not bets, but the formation of dispositions in relation to landscapes of probability.

The discourse around explainable AI (XAI) usually criticises machine learning for producing opaque models (so-called 'black boxes') and demands more transparency in algorithmic reasoning. The article *Poetics of Opacity: Glissant and Bayes* by Conrad Moriarty-Cole questions this implicit superiority of transparency over opacity in XAI debates by turning to Édouard Glissant's concept of opacity, which was developed in response and resistance to the violent imposition of transparency over enslaved people during colonialism. After arguing for the persistence of machinic opacity, the article investigates whether a 'poetics of opacity' that extends Glissant's notion of 'inter-cultural opacity' to 'intra-cultural opacity' could help us construct an 'ethicopolitics' that acknowledges this opacity in a non-reductive way. The article then turns to Bayesian inference to show that it has an ambiguous relationship with opacity: it is problematic when presented as a technical tool to render opacity as a quantity, but it also opens up possibilities for a poetics of opacity by incorporating prior knowledge of the poet, artist, activist or community.

In the article *Situating Bayesian Knowledge: A Case Study of Modelling Pollutant Transfers from Land to Water*, Krystin Unverzagt, Tobias Krueger, Anja Klein, Márk Somogyvári and Rossella Alba explore the ways that Bayesian approaches are situated and contingent. The authors look into a specific case of Bayesian modelling of land-to-water pollutant transfers and investigate how specific modeling practices enact onto-epistemic constructs, positioning such modelling as a technoscientific endeavor that, like all such endeavors, requires accounting for the kinds of connections that it draws. The authors approach the case of Bayesian modelling from an STS perspectives, particularly paying attention to the material-semiotic situatedness (Haraway) and enactments (Mol) through which knowledge objects that were initially undetermined and plural become singularized and concretised. The paper thus shifts focus from discussion of Bayesian epistemological principles and instead towards relations and embeddings within which Bayesian knowledge is enacted and the consequences thereof. The authors argue that the way Bayesian knowledge is 'done', such as through modelling and inference practices, positions it as partial. Through their case study, Unverzagt et al. meticulously trace how such enactments emerged and how they could have been different.

# 6 Situated Bayes - a research agenda

Each of the articles collected in this special issue addresses a facet of the wider question of a relation between feminist epistemology and decolonial and pluriversal relations to knowledge as articulated to Bayesian approaches in mathematics, computing, data science and fields where they are applied. Each in turn produces further elaborations and questionings of the aspects of this question that is opened here. These articles signal some of the directions for a research agenda towards more situated Bayesian approaches and their operationalisations, as well as generative critique. Such an agenda, we argue, needs to be articulated on different scales and within different domains, including: mathematics (and its adjacent approaches), informatics including machine learning and human–computer interaction (HCI), infrastructures and situated sociotechnical implementations of various concrete abstractions that these disciplines produce, as well as their philosophical and cultural dimensions. Software studies, of which this journal is a part, may operate as a means of bringing some of these together. Indeed, while these areas have their own modes of method and practice, their intersections are also important to attend to—both as loci of mutual actualisation (e.g., such as examples that bring Bayesian machine learning techniques and feminist concepts of intersectionality[101]), as well as spaces for diffractive invention.[102]

The wider question of opening the richness of mathematical culture up to inventive and critical thought and practices also needs to be coupled with those that are internal to mathematics itself, but also to strive towards the constitution of a wider and multiform field of formation in which the basic 'rules of the game' in mathematics, computing, data science and related fields might also be re-articulated and (as we suggest in the discussion of mathematics as media) open to forms of mutation from the political questions of epistemology.

Bayes theorem is appealing in this context because it builds in an equivocation about belief and data in some ways. It is a measure of a belief or the intersection of multiple beliefs, translated into a numerical quanta. Thus, it is not a belief itself, unless it is one that is already solely statistical, that is of concern, but a more or less adequate means of mapping a belief to a number that is comparable to other numbers. Émile Borel's 1924 intervention, of proposing to measure belief by the amount of money a person was willing to bet on a proposed outcome of a test, is one such example of this.[103] In his timely discussion of the usefulness of Bayesian statistics to predictive capitalism and the surveillance economies of platforms, Justin Jocque shows one way in which this connection can be even more closely bound together.[104] These provide good grounds for being suspicious of the argument we are trying to make, but not all enumerations are equal.

---

[101] James Foulds, Rashidul Islam, Kamrun Keya, and Shimei Pan, "Bayesian Modeling of Intersectional Fairness: The Variance of Bias", 2018, https://doi.org/10.48550/arXiv.1811.07255.

[102] For instance, while not Bayesian per se, diffractive space as a place for inventive methods to emerge between machine learning systems design and intersectional feminist methodologies was explored by Goda Klumbytė in the speculative design workshops, during which feminist theoretical concepts were actualised in a machine learning systems design framework. See Goda Klumbytė, Claude Draude, and Alex S. Taylor, "Critical Tools for Machine Learning: Working with Intersectional Critical Concepts in Machine Learning Systems Design", in *Proceedings of the 2022 ACM Conference on Fairness, Accountability, and Transparency, FAccT '22* (New York, NY, USA: Association for Computing Machinery, 2022), 1528–41, https://doi.org/10.1145/3531146.3533207.

[103] Émile Borel, "A propos d'un traité de probabilités", *Revue Philosophique de la France et de l'Étranger* 98 (1924): 321–336.

Bayesian approaches also offer the chance to learn even from inadequate or missing data from approximations and what is not known.[105] This includes ambiguous and indirect data, data that is the result of numerous translations, through, perhaps various layers of models of what it is describing. An historical example of this is Harold Jeffrey's work on calculating the strength and location of earthquakes in order to better predict future ones, in which aggregate data from multiple—often variably accurate—seismographs, measured the timing and strength of seismic waves arriving through variably composed and unknown geological material.[106] The subtlety and multi-layered nature of such work, as it brings together multiple factors in an emergent way, in an engagement with the unknown, through the interaction of possibilities, shows the decisive and inviting potential of such work. The term 'subjective Bayes' is used to describe this quality, that of the necessity to work from the position of an initial hunch or more or less informed guess as to the initial values of a probability distribution $p$ or the factors that inform it. In this article, we have proposed various means of conjoining the subjective with the situated.

A distinction can be made between the work of Bayes, along with that of Pierre-Simon Laplace and the notion of the pluriversal or situated in that they both compared the understanding of probability via mathematics to the omniscient knowledge of God as an all-knowing entity, composed perhaps by the universe itself. Laplace contrasted this with the 'frailty' and necessary partiality of human knowledge.[107] For Laplace, science was a means to come incrementally close to a more perfect knowledge but could never be that of God. For Richard Price, in his introduction and preparation of Bayes' manuscript, Bayes' theorem offered an argument for a 'final cause' of all things since the use of the theorem allowed for understanding 'in every case of any particular order or recurrency of events, what reason there is to think that such recurrency or order is derived from stable causes or regulations in nature, and not from any irregularities of chance.'[108] The hunch or initial guess, the subjective prior, is analogous to one's situatedness or frailty in relation to this imaginary omniscience. Indeed, numerous scientists of the era describe their work in relation to the notion of God. For some, there is a trepidation about trespassing on sacred domains with the earthly concerns of science. For others, God provides a means of describing nature without resorting to an open materialism. In either case, negotiating an epistemological rupture or paradigm shift in knowledge, God is a way of describing certain kinds of limits on the capacity of scientific knowledge to know and to calculate everything.[109] This contrasts with the contemporary 'God trick', mentioned in Section 4, Haraway's term for the omniscient view that is imagined for a certain kind of science.[110] The argument for the pluriversal argues, firstly, that knowledge, and many variant forms of it, are fully part of the formation of the universe and, secondly, that this notion of—even imaginary—omniscience is a hallucination that produces the intellectual desiring machines core to colonial thought. In contrast, both Bayes and Laplace use the image of God as a way of asserting the partiality of human knowledge. Theirs is an inverse of the contemporary God trick, and one that sustains the assertion of situated knowledge.

---

[104] Justin Joque, *Revolutionary Mathematics: Artificial intelligence, statistics and the logic of capitalism* (Verso, 2022).

[105] This is illustrated by Haas' story about a fairy that offers a bargain to two statisticians, which requires them to predict how a dice will roll after seeing just a single roll. The subjective Bayesian statistician chooses the same number as observed in the first roll and wins the prize while the frequentist statistician gets frustrated about not knowing whether the dice is fair. See Haas, "The Fabulous Five – A Bayesian Fairy Tale".

[106] See, Harold Jeffreys, *The Earth: Its Origin, History and Physical Constitution*, 6th ed. (Cambridge University Press, 1976). For historical context, see, David Howie, *Interpreting Probability: Controversies and Developments in the Early Twentieth Century* (Cambridge University Press, 2002).

[107] Pierre-Simon Laplace, "Sur le principe de la gravitation universelle et sur les inégalités séculaires des planètes qui en dependent", *Ouevres Completes* 8 (1776): 201-278.

[108] Richard Price, prefatory remarks to Bayes, "An Essay towards Solving a Problem in the Doctrine of Chances".

Mathematics, always incomplete, is a great way to think and to understand mathematically, but it is also possible to use it to break out of its tendency to tautology. (Analogously to the way diagnosed for language by Kharms and reworked by Guattari's reinvention of the abstract machine.) Returning to Whitehead, we could say that there is a political significance in the most abstract mathematical formulations, because they are a way into the most concrete material questions. This leads to a positive formulation of mathematics as a cultural and social construct. If mathematics is so constructed, what kind of mathematics do we want to construct? In short, we want to construct mathematics that is in dialogue with feminist and decolonial perspectives.

This is a challenging mission and many different ways into the question are possible. Mathematics is often embedded in and reworked by practices, not the least, those of computing. We don't have, and nor do we want, solutions for fully resolving the contradictions between Bayesian approaches and feminist and decolonial perspectives. Rather, 'Situated Bayes' sketches a transdisciplinary research agenda for the future.

## Code

Python code to reproduce the numerical experiments in this article can be found at: https://github.com/juni-schindler/situated-bayes

## Acknowledgements

JS would like to thank Armin Haas at IASS Potsdam for first introducing them to Bayesian statistics during an undergraduate research placement in 2018. We would like to thank the participants in this issue for their dedicated work on the topic, and in particular to acknowledge the many peer reviewers who provided extensive and insightful feedback. 💗 Thanks especially to the reviewers of this article.

Funding in direct support of this work: JS acknowledges support from the EPSRC (PhD studentship through the Department of Mathematics at Imperial College London).

---

[109] For the formulation of the epistemological rupture, see Gaston Bachelard, *The Formation of the Scientific Mind: A Contribution to a Psychoanalysis of Objective Knowledge*, trans. Mary McAllester Jones (Clinamen Press, 2002). For the proposal of the paradigm shift, see Thomas Kuhn, *The Structure of Scientific Revolutions* (University of Chicago Press, 1962).

[110] Haraway, "Situated Knowledges".

# Bibliography

Abril-Pla, Oriol, Virgile Andreani, Colin Carroll, Larry Dong, Christopher J. Fonnesbeck, Maxim Kochurov, Ravin Kumar, et al. "PyMC: A Modern, and Comprehensive Probabilistic Programming Framework in Python." *PeerJ Computer Science* 9 (2023): e1516. https://doi.org/10.7717/peerj-cs.1516.

Alexander, Amir. "The imperialist space of Elizabethan mathematics." *Studies in History and Philosophy of Science* 26, no. 4 (1995): 559-591.

Bachelard, Gaston. *The Formation of the Scientific Mind: A Contribution to a Psychoanalysis of Objective Knowledge*. Translated by Mary McAllester Jones. Clinamen Press, 2002.

Baker, Michael. "The Western mathematic and the ontological turn: Ethnomathematics and cosmotechnics for the pluriverse." In *Indigenous Knowledge and Ethnomathematics*, edited by Eric Vandendriessche and Rik Pinxten. Springer, 2022. 243–76.

Bal, Mieke. *Travelling Concepts in The Humanities: A Rough Guide*. University of Toronto Press, 2002.

Barad, Karen Michelle. *Meeting the Universe Halfway: Quantum Physics and the Entanglement of Matter and Meaning*. Duke University Press, 2007.

Bayes, Thomas. "An Essay towards Solving a Problem in the Doctrine of Chances. By the Late Rev. Mr. Bayes, F.R.S. Communicated by Mr. Price, in a Letter to John Canton, A.M.F.R.S." *Philosophical Transactions of the Royal Society of London* 53 (1763): 370–418.

Bhattacharya, Nandini, Regina D. Langhout, S. Sylvane Vaccarino-Ruiz, et al. "'Being a team of five strong women… we had to make an impression:' The College Math Academy as an intervention into mathematics education." *American Journal of Community Psychology* 70, no. 1-2 (2021): 228-241. https://doi.org/10.1002/ajcp.12573.

Black, Laura, Diane Harris, Tee McCaldin, Kate O'Brien, Maria Pampaka, Julian Williams, and Shunqi Zhang. "The Connection between Gender and School Mathematics: The Views of Queer, Trans and Gender Non-Conforming Students." 2025. https://assets.nicepagecdn.com/5826c5c3/6355331/files/Black_TWG10.pdf.

Blas, Zach, Melody Jue, and Jennifer Rhee, eds. *Informatics of Domination*. Duke University Press, 2025.

Bellhouse, D. R. "The Reverend Thomas Bayes, FRS: A Biography to Celebrate the Tercentenary of His Birth." *Statistical Science* 19, no. 1 (2004): 3–43.

Benjamin, Ruha. *Race After Technology: Abolitionist Tools for the New Jim Code*. Polity, 2019.

Bishop, Alan J. "Western mathematics: the secret weapon of cultural imperialism." *Race & Class* 32, no. 2 (1990): 51-65.

Borel, Émile. "A propos d'un traité de probabilités." *Revue Philosophique de la France et de l'Étranger* 98 (1924): 321–336.

Buck, Caitlin E., and Bo Meson. "On Being a Good Bayesian." *World Archaeology* 47, no. 4 (2015): 567–84. https://doi.org/10.1080/00438243.2015.1053977.

de la Cadena, Marisol. *Earth Beings: Ecologies of Practice Across Andean Worlds.* Duke University Press, 2015.

de Castro, Eduardo Viveiros. *Cannibal Metaphysics*. Translated by Peter Skafish. Univocal Press, 2014.

Chehonadskih, Maria. *Alexander Bogdanov and the Politics of Knowledge After the October Revolution*. Palgrave Macmillan, 2023.

Chronaki, Anna, and Dalene Swanson. "De/mathematising the political: bringing feminist de/post-coloniality to mathematics education." In *Quaderni di Ricerca in Didattica' QRDM (Mathematics)* 27, suppl. no. 2 (2017): 67-71.

Chun, Wendy Hui Kyong. *Discriminating Data: Correlation, Neighborhoods, and the New Politics of Recognition*. MIT Press, 2024.

Clark, Andy. *Surfing Uncertainty: Prediction, Action, and the Embodied Mind*. Oxford University Press, 2016.

Couldry, Nick, and Ulises Ali Mejias. "Data Colonialism: Rethinking Big Data's Relation to the Contemporary Subject". *Television & New Media* 20, no. 4 (2019): 336–49. https://doi.org/10.1177/1527476418796632.

Damarin, Suzanne. "Toward Thinking Feminism and Mathematics Together." *Signs: Journal of Women in Culture and Society* 34, no. 1 (2008): 101–23. https://doi.org/10.1086/588470.

Debaise, Didier. *Nature as Event: The Lure of the Possible*. Translated by Michael Halewood. Duke University Press, 2017.

Deleuze, Gilles. *Difference and Repetition*. Translated by Paul Patton. Columbia University Press, 1995.

Derrida, Jacques. *Specters of Marx: The State of the Debt, the Work of Mourning, & the New International*. Translated by Peggy Kamuf. Routledge, 1994.

D'Ignazio, Catherine, and Lauren F. Klein. *Data Feminism*. The MIT Press, 2020.

van der Drift, Mijke, and Nat Raha. *Trans-Femme Futures*. Pluto Press, 2024.

Eglash, Ron, Audrey Bennett, Casey O'Donnell, Sybillyn Jennings, and Margaret Cintorino. "Culturally Situated Design Tools: Ethnocomputing from Field Site to Classroom." *American Anthropologist* 108, no. 2 (2006): 347–62. http://www.jstor.org/stable/3804796.

Eglash, Ron. "Ethno-biomathematics: A Decolonial Approach to Mathematics at the Intersection of Human and Nonhuman Design." In *Ubiratan D'Ambrosio and Mathematics Education*, edited by Marcelo C. Borba and Daniel C. Orey. Springer, 2023. https://doi.org/10.1007/978-3-031-31293-9_18.

Escobar, Arturo. *Designs for the Pluriverse: Radical Interdependence, Autonomy, and the Making of Worlds*. Duke University Press, 2018.

Eyache, Elie. *The Blank Swan: The End of Probability*. John Wiley & Sons, 2010.

Foulds, James, Rashidul Islam, Kamrun Keya, and Shimei Pan, "Bayesian Modeling of Intersectional Fairness: The Variance of Bias", 2018, https://doi.org/10.48550/arXiv.1811.07255.

Foucault, Michel. *The Order of Things.* Taylor & Francis, 2012. http://public.eblib.com/choice/publicfullrecord.aspx?p=240649.

Fuller, Matthew, and Andrew Goffey. *Evil Media.* MIT Press, 2012.

Fuller, Matthew, and Eyal Weizman. *Investigative Aesthetics: Conflicts and Commons in the Politics of Truth*. Verso, 2021.

Fúnez-Flores, J.I., W. Ataide Pinheiro, A. Ávila Mendoza, R. Phelps and E. Cherry Shive. "The sociopolitical turn in mathematics education and decolonial theory." *London Review of Education* 22, no. 1 (2024): 13. https://doi.org/10.14324/LRE.22.1.13.

Gelman, Andrew, John B. Carlin, Hal S. Stern, David B. Dunson, Aki Vehtari, and Donald B. Rubin. *Bayesian Data Analysis*. 3rd ed. CRC Press, 2014.

Gelman, Andrew, and Cosma Rohilla Shalizi. "Philosophy and the Practice of Bayesian Statistics." *British Journal of Mathematical and Statistical Psychology* 66, no. 1 (2013):

8–38. https://doi.org/10.1111/j.2044-8317.2011.02037.x.

Ghahramani, Zoubin. "Probabilistic Machine Learning and Artificial Intelligence." *Nature* 521, no. 7553 (2015): 452–59. https://doi.org/10.1038/nature14541.

Gitelman, Lisa, ed. *"Raw Data" is an Oxymoron*. MIT Press, 2013.

Glissant, Édouard. *Poetics of Relation*. Translated by Betsy Wing. University of Michigan Press, 1997.

Guattari, Félix. *Chaosmosis: An Ethico-Aesthetic Paradigm*. Translated by P. Bains and J. Pefanis. Power Publications, 1995.

Haas, Armin. "The Fabulous Five – A Bayesian Fairy Tale." In *Textures of the Anthropocene*, 209–22. Cambridge, MA: MIT Press, 2015. https://publications.rifs-potsdam.de/pubman/faces/ViewItemFullPage.jsp?itemId=item_2114891_4.

Hacking, Ian. *Logic of Statistical Inference*. Cambridge University Press, 2016.

O'Hagan, Anthony, Caitlin E. Buck, Alireza Daneshkhah, J. Richard Eiser, Paul H. Garthwaite, David J. Jenkinson, Jeremy E. Oakley, and Tim Rakow. *Uncertain Judgements: Eliciting Experts' Probabilities*. Chichester: John Wiley & Sons, Ltd, 2006.

Haraway, Donna. "Situated Knowledges: The Science Question in Feminism and the Privilege of Partial Perspective." *Feminist Studies* 14, no. 3 (1988): 575. https://doi.org/10.2307/3178066.

Harding, Sandra. "Postcolonial and feminist philosophies of science and technology: convergences and dissonances." *Postcolonial Studies* 12, no. 4 (2009): 401-421.

Harding, Sandra. "Rethinking Standpoint Epistemology: What is 'Strong Objectivity?" In *Feminist Theory: A Philosophical Anthology*, edited by Ann E. Cudd and Robin O. Andreasen. Oxford: Blackwell, 2005.

Harding, Sandra G. *The Science Question in Feminism*. Cornell University Press, 1986.

Hardy, Godfrey H. *A Mathematician's Apology*. 20th print. Canto Classics. Cambridge University Press, 2013.

Hastie, Trevor, Robert Tibshirani, and Jerome Friedman. *The Elements of Statistical Learning*. Springer, 2009.

Hill Collins, Patricia. *Black Feminist Thought*. Routledge, 1990.

Hoffman, Matthew D., and Andrew Gelman. "The No-U-Turn Sampler: Adaptively Setting Path Lengths in Hamiltonian Monte Carlo." *Journal of Machine Learning Research* 15, no. 47 (2014): 1593–1623.

Hornik, Kurt, Maxwell Stinchcombe, and Halbert White. "Multilayer Feedforward Networks Are Universal Approximators." *Neural Networks* 2, no. 5 (1989): 359–66. https://doi.org/10.1016/0893-6080(89)90020-8.

Hottinger, Sara N. *Inventing the Mathematician: Gender, Race and Our Cultural Understanding of Mathematics*. State University of New York Press, 2016.

Howie, David. *Interpreting Probability: Controversies and Developments in the Early Twentieth Century*. Cambridge University Press, 2002.

Hui, Yuk. *Machine and Sovereignty: For a Planetary Thinking*. University of Minnesota Press, 2024.

Hui, Yuk. *Recursivity and Contingency*. Rowman and Littlefield International, 2019.

Joque, Justin. *Revolutionary Mathematics: Artificial Intelligence, Statistics and the Logic of Capitalism*. Verso, 2022.

Jablonka, Eva. "Mathematisation in environments of Big Data – 'implicit mathematics' revisited." In *Proceedings from CIEAEM 69, Chapter 1: Plenaries*, edited by B. di Paola and U. Gellert, 43-51. *Quaderni di Ricerca in Didattica (Mathematics)* 27, suppl. no. 2, 2017. https://sites.unipa.it/grim/quaderno27_suppl_2.htm.

Jeffreys, Harold. *The Earth: Its Origin, History and Physical Constitution*. 6th ed. Cambridge

University Press, 1976.

Kenney, Martha. "Counting, Accounting and Accountability: Helen Verran's Relational Empiricism." *Social Studies of Science* 45, no. 5 (2015): 749-771.

Khan, S., S. LaFrance, and H. T. T. Tran. "After plantations' precarities: curating math-thematic curriculum plots in initial teacher education for multispecies' flourishing and a freedom-yet-to-come." *Research in Mathematics Education* 24, no. 2 (2022): 170–186. https://doi.org/10.1080/14794802.2022.2090421.

Kharms, Daniil. "Blue Notebook no.10." In *Today I wrote Nothing: The Selected Writings of Daniil Kharms*. Translated by Matvei Yankelevich, p.45. Ardis, 2009.

Kitchin, Rob. *The Data Revolution: Big Data, Open Data, Data Infrastructures & Their Consequences*. SAGE Publications, 2014.

Klumbytė, Goda, Claude Draude, and Alex S. Taylor. "Critical Tools for Machine Learning: Working with Intersectional Critical Concepts in Machine Learning Systems Design." In *Proceedings of the 2022 ACM Conference on Fairness, Accountability, and Transparency*, FAccT '22, 1528–41. New York, NY, USA: Association for Computing Machinery, 2022. https://doi.org/10.1145/3531146.3533207.

Laplace, Pierre-Simon. "Sur le principe de la gravitation universelle et sur les inégalités séculaires des planètes qui en dependent." *Ouevres Completes* 8 (1776): 201-278.

Law, John. "What's Wrong with the One World World?" *Distinktion: Scandinavian Journal of Social Theory* 16, no. 1 (2015): 126–39.

Loukissas, Yanni. *All Data are Local: Thinking Critically in a Data-Driven Society*. MIT Press, 2019.

Lugones, Maria. "Toward a Decolonial Feminism." *Hypatia* 25, no. 4 (2010): 742-759. https://www.jstor.org/stable/40928654.

Kuhn, Thomas. *The Structure of Scientific Revolutions*. University of Chicago Press, 1962.

Lynch, Scott M., and Bryce Bartlett. "Bayesian Statistics in Sociology: Past, Present, and Future." *Annual Review of Sociology* 45, no. 1 (2019): 47–68. https://doi.org/10.1146/annurev-soc-073018-022457.

Mackenzie, Adrian. *Machine Learners: Archaeology of a Data Practice*. The MIT Press, 2017.

McGuire, Linda. "Feminist Theories Informing Mathematical Practice." In *Handbook of the Mathematics of the Arts and Sciences*, edited by B. Sriraman. Cham: Springer, 2020. https://doi.org/10.1007/978-3-319-70658-0_77-1.

McGrayne, Sharon Bertsch. *The Theory That Would Not Die.* Yale University Press, 2011.

McKittrick, Katherine. "Mathematics Black Life." *The Black Scholar* 44, no. 2 (2014): 16-28.

Meillasoux, Quentin. *After Finitude: An Essay on the Necessity of Contingency*. Translated by Ray Brassier. Continuum, 2008.

Mignolo, Walter. "Delinking - The Rhetoric of Modernity, the Logic of Coloniality and the Grammar of de-Coloniality." *Cultural Studies* 21, no. 2–3 (2007): 449–514. https://doi.org/10.1080/09502380601162647.

Mignolo, Walter. *Local Histories/Global Designs: Coloniality, Subaltern Knowledges, and Border Thinking*. Princeton University Press, 2000.

Mignolo, Walter. *The Darker Side of Western Modernity: Global Futures, Decolonial Options*. Duke University Press, 2011.

Morey, Richard D., Jan-Willem Romeijn, and Jeffrey N. Rouder. "The Philosophy of Bayes Factors and the Quantification of Statistical Evidence." *Journal of Mathematical Psychology* 72 (2016): 6–18. https://doi.org/10.1016/j.jmp.2015.11.001.

Nelson, Diane M. *Who Counts?: The Mathematics of Death and Life after Genocide*. Duke University Press, 2015.

Nietzsche, Friedrich. *On the Genealogy of Morality*. Translated by Carol Diethe. Edited by Keith Ansell-Pearson. Cambridge University Press, 2017.

Noble, Safiya Umoja. *Algorithms of Oppression: How Search Engines Reinforce Racism*. New York University Press, 2018.

Quijano, Aníbal. "Coloniality and Modernity/Rationality." *Cultural Studies* 21, no. 2–3 (2007): 168–178. https://doi.org/10.1080/09502380601164353.

Rasmussen, Carl Edward. "Gaussian Processes in Machine Learning." In *Advanced Lectures on Machine Learning: ML Summer Schools 2003, Canberra, Australia, February 2 - 14, 2003, Tübingen, Germany, August 4 - 16, 2003, Revised Lectures*, edited by Olivier Bousquet, Ulrike von Luxburg, and Gunnar Rätsch, 63–71. Berlin, Heidelberg: Springer, 2004. https://doi.org/10.1007/978-3-540-28650-9_4.

Rotman, Brian. *Signifying Nothing: The Semiotics of Zero*. Stanford University Press, 1993.

Roux, Kate le, and Dalene Swanson. "Toward a Reflexive Mathematics Education within Local and Global Relations: Thinking from Critical Scholarship on Mathematics Education within the Sociopolitical, Global Citizenship Education and Decoloniality." *Research in Mathematics Education* 23, no. 3 (2021): 323–37. https://doi.org/10.1080/14794802.2021.1993978.

Shulman, Bonnie. "What if We Change Our Axioms? A Feminist Inquiry into the Foundations of Mathematics." *Configurations* 4, no. 3 (1996): 449. https://doi.org/10.1353/con.1996.0022.

Simondon, Gilbert. *Imagination and Invention*. Translated by Joe Hughes & Christophe Wall-Romana. Minnesota University Press, 2023.

Schindler, Juni, and Matthew Fuller. "Community as a Vague Operator: Epistemological Questions for a Critical Heuristics of Community Detection Algorithms." *Computational Culture* 9 (2023). http://computationalculture.net/community-as-vague-operator/.

Serres, Michel. *Hermes 1: Communication*. Translated by Louise Burchill. University of Minnesota Press, 2023

Sohn-Rethel, Alfred. *Intellectual and Manual Labour: A Critique of Epistemology.* Translated by Martin Sohn-Rethel. Haymarket Books, 2021.

Spivak, Gayatri Chakravorty. "Can the Subaltern Speak?" In *Marxism and the Interpretation of Culture*, 271–313. London: Macmillan Education, 1988.

Sprenger, Jan, and Stephan Hartmann. *Bayesian Philosophy of Science*. 1st ed. Oxford University Press, 2019.

Stengers, Isabelle. "Introductory Notes on an Ecology of Practices." *Cultural Studies Review* 11, no 1 (2005): 183-196.

Vanderschiesse, Eric, and Rik Pinxten (eds.). *Indigenous Knowledge and Ethnomathematics*. Springer, 2022.

Vapnik, Vladimir, and Rauf Izmailov. "Complete Statistical Theory of Learning: Learning Using Statistical Invariants." *Proceedings of the Ninth Symposium on Conformal and Probabilistic Prediction and Applications (Conformal and Probabilistic Prediction and Applications, PMLR)* (2020): 4–40. https://proceedings.mlr.press/v128/vapnik20a.html.

Verran, Helen. *Science and an African Logic*. University of Chicago Press, 2001.

Tenen, Dennis. *Plain Text: The Poetics of Computation*. Stanford University Press, 2017.

Whitehead, Alfred North. *Science and the Modern World: Lowell Lectures, 1925*. The Free Press, 1997.

Zalamea, Fernando. *Synthetic Philosophy of Contemporary Mathematics*. Urbanomic, 2014.